\documentclass[journal]{IEEEtran}

\ifCLASSINFOpdf
\else
\fi

\usepackage{setspace}
\usepackage{bbm}
\usepackage[cmex10]{amsmath}
\usepackage{amssymb}
\usepackage{cite}
\usepackage{graphicx}
\usepackage{array,color}
\usepackage{amsmath}
\allowdisplaybreaks
\usepackage{stfloats}
\usepackage{graphicx}
\usepackage{epstopdf}
\usepackage{subfigure}
\usepackage{tabularx}
\usepackage{epsfig,epsf,color,balance,cite}
\usepackage{algorithmic}
\usepackage{algorithm}
\usepackage{url}
\usepackage{bm}
\usepackage{multirow}

\usepackage{amsthm}

\ifodd 0
\usepackage{soul}
\usepackage{color}
\setstcolor{red}

\newcommand{\rev}[1]{{\color{red}#1}} 
\newcommand{\del}[1]{\st{#1}} 

\newcommand{\com}[1]{\textbf{\color{red} (COMMENT: #1)}} 
\newcommand{\response}[1]{\textbf{\color{green} (RESPONSE: #1)}} 
\else

\newcommand{\rev}[1]{#1}

\newcommand{\del}[1]{}

\newcommand{\com}[1]{}
\newcommand{\comg}[1]{}
\newcommand{\response}[1]{}
\fi

\title{\huge {Intelligent Reflecting Surface Assisted Multi-User OFDMA: Channel Estimation and Training Design}}

\author{ Beixiong Zheng,~\IEEEmembership{Member,~IEEE}, Changsheng You,~\IEEEmembership{Member,~IEEE}, 
	and Rui Zhang,~\IEEEmembership{Fellow,~IEEE} 
	\thanks{
		
		The authors are with the Department of Electrical and Computer Engineering, National University of Singapore, Singapore 117583,
		email: \{elezbe, eleyouc, elezhang\}@nus.edu.sg.

	}
}

\begin{document}
\markboth{IEEE Transactions on Wireless Communications, Vol. XX, No. XX, XXX 2020}{SKM: My IEEE article}
\maketitle
\vspace{-1cm}
\begin{abstract}
	To achieve the full passive beamforming gains of intelligent reflecting surface (IRS), accurate
	channel state information (CSI) is indispensable but practically challenging to acquire,
	due to the excessive amount of channel parameters to be estimated which increases with the number of IRS reflecting elements as well as that of IRS-served users.
	To tackle this challenge, we propose in this paper two efficient channel estimation schemes for different channel setups in an IRS-assisted multi-user broadband communication system employing the orthogonal frequency division multiple access (OFDMA).
	The first channel estimation scheme, which estimates the CSI of all users in parallel simultaneously at the access point (AP), is applicable for arbitrary frequency-selective fading channels. In contrast, the second channel estimation scheme, which exploits a key property that all users share the same (common) IRS-AP channel to enhance the training efficiency and support more users, is proposed for the typical scenario with line-of-sight (LoS) dominant user-IRS channels.
	For the two proposed channel estimation schemes,
	we further optimize their corresponding training designs (including pilot tone allocations for all users and IRS time-varying reflection pattern) to minimize the channel estimation error. Moreover, we 
	derive and compare the fundamental limits on the minimum training overhead and the maximum number of supportable users of these two schemes.
	 Simulation results verify the effectiveness of the proposed channel estimation schemes and training designs, and 
	show their significant performance improvement over various benchmark schemes.

\end{abstract}
\IEEEpeerreviewmaketitle

\section{Introduction}
\IEEEPARstart{D}{riven}
 by the skyrocketing growth of mobile devices and wide deployment of Internet of things (IoT), various advanced wireless technologies such as massive multiple-input multiple-output (MIMO), millimeter wave (mmWave) and network densification, have been proposed and extensively investigated in the last decade for substantially improving the network capacity and connectivity of wireless communication systems \cite{Andrews2014What}.
However, the performance improvement of these technologies generally comes at the expense of increased network energy consumption and hardware complexity due to the ever-increasing number of active antennas/radio-frequency (RF) chains, which incurs high system implementation cost and may hinder their future applications.
Moreover, due to the lack of control over the wireless propagation channel, these technologies need to adapt to the time-varying wireless environments, which, however, cannot always guarantee the quality-of-service (QoS) with uninterrupted connectivity in some harsh propagation conditions (e.g., severe attenuation and poor diffraction due to the blockage of wireless communication links in mmWave frequency bands).

Leveraging the recent advances in reconfigurable meta-surfaces \cite{cui2014coding,liaskos2018new,liu2018programmable}, intelligent reflecting surface (IRS) (a.k.a. reconfigurable intelligent surface or other equivalents) has emerged as an innovative technology to achieve cost-effective improvement in communication coverage, throughput, and energy efficiency \cite{qingqing2019towards,Renzo2019Smart,basar2019wireless,Wu2019TWC,wu2019beamforming}.
Different from the existing technologies that are only able to adapt to the dynamic wireless channels,
IRS can program the signal propagation by intelligently controlling a large number of
passive reflecting elements (e.g., low-cost printed dipoles \cite{yang2016design}), each of which is capable of altering the amplitude and/or phase of the reflected signal, thus collaboratively enabling the real-time reconfiguration of wireless propagation environment.
Furthermore, IRS can achieve full-duplex passive beamforming without requiring any costly processing for self-interference cancellation and signal decoding/amplification, thus substantially reducing the complexity, energy consumption, and hardware cost.
\rev{These appealing advantages have motivated active research on the joint design of IRS with other wireless techniques, e.g., orthogonal frequency division multiplexing (OFDM) \cite{zheng2019intelligent,yang2019intelligent,Yang2020IRS}, 
multi-antenna communication \cite{zhang2019capacity,Pan2020Multicell}, 
non-orthogonal multiple access (NOMA) \cite{Zheng2020IRSNOMA,yanggang2019intelligent},
 wireless power transfer \cite{Wu2020Weighted,Wu2020Weighted2,Pan2020Intelligent},
physical layer security \cite{Cui2019Secure,Guan2020Intelligent,chen2019intelligent}, 
cognitive radio \cite{Guan2020Joint},
and so on.}

To fully achieve the passive beamforming gains of IRS, the acquisition of accurate channel state information (CSI) at the access point (AP)/IRS is of paramount importance in practice, which, however, is fundamentally challenging due to the following reasons. 
First, without any active components, the passive IRS elements are lack of baseband processing capabilities and thus incapable of transmitting/receiving pilot signals, which makes the conventional pilot-aided channel estimation by IRS inapplicable.
As such, an alternative approach is to estimate the cascaded user-IRS-AP channels at the AP based on the user pilot signals and time-varying IRS reflection pattern \cite{zheng2019intelligent,yang2019intelligent}.
Second, due to the massive number of IRS reflecting elements, it is practically difficult to
estimate the full CSI associated with each reflecting element given a limited channel training time.
\rev{To reduce the training overhead with the increasing number of IRS elements as well as simplify the passive beamforming design,
a novel elements-grouping strategy was proposed in \cite{zheng2019intelligent,yang2019intelligent}, which groups adjacent IRS elements (typically with high channel correlation) into a sub-surface and thus only needs to estimate the effective cascaded user-IRS-AP channel associated with each sub-surface.
Moreover, a flexible system trade-off between training overhead and passive beamforming performance can be achieved by adjusting the size of each sub-surface \cite{zheng2019intelligent,yang2019intelligent}.}
Furthermore, to improve the channel estimation accuracy of the ON/OFF-based IRS reflection pattern design that does not fully exploit the IRS array gain \cite{yang2019intelligent}, a discrete Fourier transform (DFT)-based IRS reflection pattern design was proposed in \cite{zheng2019intelligent,jensen2019optimal} to achieve the minimum channel estimation error. 
\rev{In \cite{you2019intelligent} and \cite{you2019progressive},
a practical IRS reflection design was proposed under the more realistic setting with discrete phase shifts \cite{wu2019beamforming} at the IRS and variable-length pilot symbols for channel training.
Moreover, the robust passive beamforming designs based on the imperfect CSI with correlated channel estimation errors were investigated in \cite{you2019intelligent,you2019progressive,zhou2020framework,zhao2019intelligent}.}
Besides, for IRS-assisted MIMO systems, various channel estimation methods have been proposed in \cite{he2019cascaded,mirza2019channel,wang2019compressed} by exploiting certain IRS channel properties such as low-rank, sparsity, spatial correlation, etc.

Note that the above-mentioned works mainly focus on channel estimation for the IRS-assisted single-user system, which, however, cannot be efficiently applied to the IRS-assisted multi-user system since the straightforward user-by-user successive channel estimation will incur prohibitive training overhead that scales with the number of users and thus may be unaffordable given a finite channel coherence time in practice. 
Although some initial channel estimation studies have been recently pursued for the IRS-assisted multi-user narrowband system \cite{chen2019channel,liu2019matrix,wang2019channel,wei2020parallel}, the fundamental limits of the multi-user channel estimation in terms of training overhead, number of supportable users as well as channel estimation performance have not been fully characterized yet, to the best of our knowledge.
Moreover, for broadband communications over frequency-selective fading channels in general, the above-mentioned multi-user channel estimation methods tailored for narrowband communications become inapplicable due to the frequency-selective fading channels but frequency-flat IRS reflections \cite{zheng2019intelligent,yang2019intelligent}, which thus calls for innovative solutions to tackle these new challenges.


\begin{figure}[!t]
	\centering
	\includegraphics[width=2in]{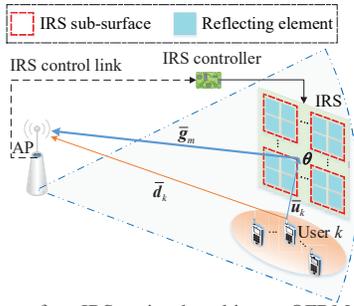}
	\setlength{\abovecaptionskip}{-6pt}
	\caption{Illustration of an IRS-assisted multi-user OFDMA uplink communication system.}
	\label{system}
	\vspace{-0.5cm}
\end{figure}

Motivated by the above, in this paper, we consider an IRS-assisted multi-user system employing the orthogonal frequency division multiple access (OFDMA), where an IRS is deployed to aid the communications between an AP and multiple users in its service region under frequency-selective fading channels, as shown in Fig.~\ref{system}.
In particular, we propose two efficient channel estimation schemes for different channel setups in the users' uplink transmissions, while the estimated CSI at the AP can also be applied to the downlink if the uplink-downlink channel reciprocity is assumed. For both schemes, we optimize their corresponding channel training designs (including the pilot tone allocations for all users and IRS time-varying reflection pattern) to minimize the channel estimation error, and characterize the minimum training overhead as well as the maximum number of supportable users. 
The main contributions of this paper are summarized as follows.
\begin{itemize}
	\item First, we consider a general IRS-assisted multi-user OFDMA system under arbitrary frequency-selective fading channels for all the involved user-AP, user-IRS, and IRS-AP links and propose a low-complexity channel estimation scheme, called \emph{simultaneous-user} channel estimation, to estimate the CSI of all users in parallel simultaneously at the AP. To unveil the fundamental limits of this scheme, we derive its minimum training time and maximum number of supportable users. Moreover, we optimize the training design in terms of user pilot tone allocations and IRS reflection pattern to minimize the channel estimation error, for which the optimal solution is derived in closed-form.
	\item Next, we consider a typical scenario where the user-IRS channels are line-of-sight (LoS) dominant and propose a new customized channel estimation scheme, called \emph{sequential-user} channel estimation, to increase the maximum number of supportable users by exploiting a key property that all the users share the same (common) IRS-AP channel.
	Specifically, the proposed new scheme first estimates the CSI of an arbitrarily-selected reference user, based on which the CSI of the remaining non-reference users is then recovered by only estimating their effective user-IRS channels normalized by that of the reference user. 
	Moreover, as the corresponding joint training design for all users is highly challenging in this case, we propose an efficient training design that first optimizes the pilot tone allocations and IRS reflection pattern for the reference user, and then solves the pilot tone allocation problem jointly for the remaining non-reference users with fixed IRS reflection pattern for the reference user.
	\item Finally, we compare the two proposed channel estimation schemes in terms of complexity, maximum number of supportable users, and minimum training overhead of each user, given the same channel training time. 
	In general, as compared to the simultaneous-user channel estimation,
	the sequential-user channel estimation is able to support more users at the expense of higher channel estimation complexity and some degraded channel estimation performance for the users.
	Moreover, we provide extensive numerical results to validate the performance improvement of our proposed 
	training designs over other benchmark schemes with different pilot tone allocations and IRS reflection patterns. 
\end{itemize}

The rest of this paper is organized as follows. Section~\ref{sys} presents the system model for the IRS-assisted multi-user OFDMA system. In Sections~\ref{Concurrent-User} and \ref{Reference-User}, we propose two channel estimation schemes for different channel setups, respectively. 
Simulation results and discussions are presented in Section \ref{Sim}. Finally, conclusions are drawn in Section \ref{conlusion}.

\emph{Notation:} 
	Upper-case and lower-case boldface letters denote matrices and column vectors, respectively.
	Upper-case calligraphic letters (e.g., $\cal{J}$) denote discrete and finite sets.
	Superscripts ${\left(\cdot\right)}^{T}$, ${\left(\cdot\right)}^{H}$, and ${\left(\cdot\right)}^{-1}$ stand for the transpose, Hermitian transpose, and matrix inversion operations, respectively.
	${\mathbb C}^{a\times b}$ denotes the space of ${a\times b}$ complex-valued matrices.
	$n \mod{a}$ denotes the modulo operation which returns the remainder after division of $n$ by $a$.
	$|\cdot|$ denotes the absolute value if applied to a complex number or the cardinality if applied to a set.
	$\lVert\cdot\rVert$ denotes the $\ell_2 $-norm,
	$\left\|  \cdot \right\|_F$ denotes the Frobenius norm,
	${\cal O}(\cdot)$ denotes the standard big-O notation,
	$\lfloor \cdot \rfloor$ is the floor function,
	$\ast$ denotes the convolution operation,
	and ${\mathbb E}\{\cdot\}$ stands for the statistical expectation.
	${\bm I}_a$, ${\bm 1}_{a\times b}$, and ${\bm 0}_{a\times b}$ denote an identity matrix of size $a \times a $, an all-one matrix of size $a \times b $, and an all-zero matrix of size $a \times b $, respectively.
	$\mathrm{diag} (\bm{x})$ returns a diagonal matrix with the elements in $\bm{x}$ on its main diagonal. 
	The relative complement of set $\mathcal{A}$ in set $\mathcal{B}$ is denoted by $\mathcal{B} \setminus \mathcal{A}$, while the union and intersection of two sets $\mathcal{A}$ and $\mathcal{B}$ are denoted by $\mathcal{A}\bigcup\mathcal{B}$ and $\mathcal{A}\bigcap\mathcal{B}$, respectively.
	The distribution of a circularly symmetric complex Gaussian (CSCG) random vector with mean vector $\bm \mu$ and covariance matrix ${\bm \Sigma}$ is denoted by ${\mathcal N_c }({\bm \mu}, {\bm \Sigma} )$; and $\sim$ stands for ``distributed as".

\vspace{-0.3cm}
\section{System Model and Problem Description}\label{sys}
As illustrated in Fig.~\ref{system},
we consider an IRS-assisted multi-user broadband wireless communication system in the uplink employing OFDMA, \rev{where an IRS 
is deployed at the cell/network edge as a dedicated helper to assist the communications between a single-antenna AP\footnote{The proposed channel estimation schemes in this paper can be readily extended to the AP with multiple antennas by estimating their associated channels in parallel.} and $K$ single-antenna
edge users, which are far from the AP\footnote{\rev{In this case, the direct AP-user links may suffer from severe path loss and blockage due to the long propagation distance between the AP and edge users.}} but in the vicinity of the IRS, denoted by the set ${\cal K} \triangleq \{1,2,\ldots, K\}$.}
By adopting a similar elements-grouping strategy as in \cite{zheng2019intelligent,yang2019intelligent}, the IRS composed of ${M}_0$ reflecting elements
is divided into $M$ sub-surfaces, denoted by the set ${\cal M} \triangleq \{1,2,\ldots, M\}$, each of which consists of $\eta= {M}_0/M$ (assumed to be an integer) adjacent elements that share a common reflection coefficient for reducing the channel estimation and passive beamforming complexity.
Moreover, the IRS is connected to a smart controller that dynamically adjusts the desired reflections of IRS elements and also exchanges (control) information with the AP via a separate wireless link \cite{qingqing2019towards,Wu2019TWC}.
In this paper, the quasi-static block fading channel model is assumed for all the involved links, which remain approximately constant within the channel coherence time.
This is a valid assumption as IRS remains at a fixed location once deployed and is practically used to mainly support low-mobility users in its neighborhood. 



\rev{In practice, since IRS is typically deployed at the cell/network edge to serve its nearby users that are far from the AP,} the user-AP and IRS-AP links usually have relatively large multi-path delay spread due to the long propagation distances and thus are modeled as frequency-selective fading channels, whereas 
the user-IRS link is also modeled generally as a frequency-selective fading channel but with much smaller multi-path delay spread, or even a frequency-flat fading channel with one (equivalent) single path due to the much shorter distances between the IRS and its served users.
\rev{Let $L_d$, $L_1$, and $L_2$ denote the maximum multi-path delay spreads (normalized by $1/B$ with $B$ denoting the system bandwidth)  of the user-AP, IRS-AP, and user-IRS links among all the users\footnote{\rev{Although different users may have different multi-path delay spreads with the AP/IRS, we take the maximum delay spread in the channel modeling without loss of generality.}}, respectively, where we have $1\le L_2 < \min\{L_1, L_d\}$.}
Accordingly, the baseband equivalent channels from user $k$ to the AP, from the IRS to the AP, and from user $k$ to the IRS are denoted by ${\bar {\bm d}}_k\in {\mathbb{C}^{L_d\times 1}}$, ${\bar {\bm G}}\triangleq [{\bar {\bm g}}_1,{\bar {\bm g}}_2,\ldots,{\bar {\bm g}}_M] \in {\mathbb{C}^{L_1\times M }} $, and ${\bar {\bm U}}_k\triangleq [{\bar {\bm u}}_{k,1},{\bar {\bm u}}_{k,2},\ldots,{\bar {\bm u}}_{k,M}]\in {\mathbb{C}^{L_2\times M}}$, respectively, where ${\bar {\bm g}}_m \in {\mathbb{C}^{L_1\times 1 }}$ and 
	${\bar {\bm u}}_{k,m} \in {\mathbb{C}^{L_2\times 1 }}$ denote the corresponding time-domain channels from sub-surface $m$ to the AP and from user $k$ to sub-surface $m$, respectively. 
Let ${\bm \theta}\triangleq[{\theta_1},  {\theta_2},\ldots,{\theta_M}]^T=[\beta_1 e^{j \phi_1}, \beta_2 e^{j \phi_2},\ldots,\beta_M e^{j \phi_M}]^T$ denote the equivalent reflection coefficients of the IRS sub-surfaces, where $\phi_m \in [0, 2\pi)$ and $\beta_m \in [0, 1]$ are
the phase shift and reflection amplitude of the $m$-th sub-surface, respectively.
To maximize the signal power reflected by the IRS and reduce the hardware cost,
we set $\beta_m=1, \forall  m\in {\cal M} $ and only consider the phase-shift design of the IRS.
Thus, the effective time-domain reflecting channel from 
user $k$ to the AP via each sub-surface $m$ can be expressed as the convolution of the user-IRS channel, the IRS reflection coefficient,
and the IRS-AP channel, which is given by
\vspace{-0.1cm}
\begin{align}\label{conv_ch}
{\bar {\bm u}}_{k,m} \ast \theta_m \ast {\bar {\bm g}}_m =\theta_m {\bar {\bm u}}_{k,m} \ast {\bar {\bm g}}_m
=\theta_m {\bar {\bm q}}_{k,m}
\end{align}
where ${\bar {\bm q}}_{k,m} \triangleq {\bar {\bm u}}_{k,m} \ast {\bar {\bm g}}_m \in {\mathbb{C}^{L_r\times 1 }} $ denotes the cascaded user-IRS-AP channel (without the effect of phase shifts) associated with each sub-surface $m$ and $L_r=L_1+L_2-1$ is the maximum delay spread of the cascaded user-IRS-AP channel.
Let $L=\max\{L_r, L_d \}$ denote the maximum delay spread of the effective time-domain channel between the users and AP, while letting ${\bm q}_{k,m}$ and ${\bm d}_k$ denote the 
zero-padded cascaded user-IRS-AP (reflecting) channel of ${\bar {\bm q}}_{k,m}$ and zero-padded user-AP (direct) channel of ${\bar {\bm d}}_k$ for user $k$, with the zero padding lengths of $L-L_r$ and $L- L_d$, respectively.
As a result, the superimposed channel impulse response (CIR) from user $k$ to the AP by
combining the user-AP (direct) channel and the cascaded user-IRS-AP (reflecting) channel in the time domain, denoted by ${\bm h}_k \in {\mathbb{C}^{L\times 1 }}$, is obtained as
\vspace{-0.1cm}
\begin{align}\label{superposed}
{\bm h}_k={\bm Q}_k {\bm \theta}+{\bm d}_k
\end{align}
where ${\bm Q}_k=[{\bm q}_{k,1},{\bm q}_{k,2},\ldots,{\bm q}_{k,M}] \in {\mathbb{C}^{L\times M }}$ denotes
the zero-padded cascaded user-IRS-AP channel matrix (without the effect of phase shifts) by stacking ${\bm q}_{k,m}$ with $m=1,\ldots,M$.
According to \eqref{superposed}, it is sufficient to estimate the cascaded reflecting channels $\{{\bm Q}_k\}_{k=1}^K$ and the direct channels $\{{\bm d}_k\}_{k=1}^K$ for the multi-user passive beamforming design in the IRS-assisted OFDMA communication system \cite{Yang2020IRS}.

For the OFDMA-based broadband communication system, the total bandwidth $B$ is equally divided into $N$ sub-carriers, which are indexed by $n\in {\cal N} \triangleq \{0,1,\ldots, N-1\}$ and
shared by the $K$ users with $N\ge K$ in general. 
Since the IRS elements have no transmit/receive RF chains, we consider the uplink training for the multi-user channel estimation at the AP over $\tau$ consecutive OFDM symbols during the time slots $t \in {\cal T}\triangleq \{1,2,\ldots,\tau\}$ of each channel coherence time.
To avoid inter-user interference and simplify the training design, we consider the disjoint pilot tone allocations for all the users in this paper,
where each sub-carrier at each time slot is allocated to at most one user.
Specifically, let $\delta_{k,n}^{(t)}$ indicate whether sub-carrier $n$ is allocated to user $k$ at time slot $t$, i.e., $\delta_{k,n}^{(t)}=1$ if sub-carrier $n$ is assigned to user $k$ at time slot $t$, and $\delta_{k,n}^{(t)}=0 $ otherwise. 
Thus, we have $\delta_{k,n}^{(t)} \in\{0,1\}$ and $\sum_{k=1}^K \delta_{k,n}^{(t)} \le 1, \forall t \in {\cal T},  \forall n \in {\cal N}$.
Here we denote ${\cal J}_{k}^{(t)}$ as the index set of the pilot tones assigned to user $k$ at time slot $t$, which is given by ${\cal J}_{k}^{(t)}\triangleq \left\{n | \delta_{k,n}^{(t)}=1 \right\}$.
As the CSI is unknown \emph{a priori}, we consider the equal transmit power allocation for each user over the assigned $|{\cal J}_{k}^{(t)}|$ sub-carriers at each time slot $t$, where the transmit power of user $k$ on each assigned sub-carrier is given by $P/|{\cal J}_{k}^{(t)}|, \forall k\in {\cal K}, \forall t \in {\cal T}$.
Let ${\bm x}_k^{(t)}\triangleq  \left[X_{k,0}^{(t)},X_{k,1}^{(t)},\ldots, X_{k,N-1}^{(t)}  \right]^T$
denote the transmitted OFDM symbol of user $k$ at time slot $t$, with each element given by
\vspace{-0.2cm}
\begin{align}\label{OFDM_sym}
X_{k,n}^{(t)}=\sqrt{\frac{P}{|{\cal J}_{k}^{(t)}|}} \delta_{k,n}^{(t)} S_{k,n}^{(t)} ,  \quad \forall t \in {\cal T},  \forall n \in {\cal N}, \forall k \in {\cal K}
\end{align}
where $S_{k,n}^{(t)}$ denotes the pilot symbol which is simply set as $S_{k,n}^{(t)}=1$ for ease of exposition, and we have $\left\|{\bm x}_k^{(t)}\right\|^2=P$.
Before transmission, each OFDM symbol ${\bm x}_k^{(t)}$ is first transformed into the time domain via an $N$-point inverse DFT (IDFT), and then appended by a cyclic prefix (CP) of length $L_{cp}$ to mitigate the inter-symbol-interference (ISI), which is assumed to satisfy $L_{cp}\ge L-1$. 
After removing the CP and performing an $N$-point DFT at the AP side,  the equivalent baseband received signal in the frequency domain is given by 
\begin{align}\label{receive}
{\bm y}^{(t)}&= \sum_{k=1}^{K} {\bm X}_k^{(t)} {\bm F} {\bm h}_k^{(t)}+ {\bm v}^{(t)}
\end{align}
where 
${\bm y}^{(t)}\triangleq \left[Y^{(t)}_{0},Y^{(t)}_{1},\ldots, Y^{(t)}_{N-1}   \right]^T$ is the received OFDM symbol at time slot $t$,
${\bm X}_k^{(t)}=\text{diag} \left({\bm x}_k^{(t)}\right)$ is the diagonal matrix of the OFDM symbol ${\bm x}_k^{(t)}$,
${\bm F}$ is an $N \times L$ matrix consisting of the $N$ rows and the first $L$ columns of the $N \times N$ unitary DFT matrix,
and ${\bm v}^{(t)}\triangleq \left[V_{0}^{(t)},V_{1}^{(t)},\ldots,V_{N-1}^{(t)}   \right]^T \sim {\mathcal N_c }({\bm 0}, \sigma^2{\bm I}_N )$ is the additive white Gaussian noise (AWGN) vector at the AP with $\sigma^2$ being the noise power. Note that the effective channel ${\bm h}_k^{(t)}$ in \eqref{receive} is time-varying over $t$ in general with dynamically tuned IRS reflection coefficients ${\bm \theta}$ over different time slots to facilitate the channel estimation (as will be shown later in this paper). As such, 
by denoting ${\bm \theta}^{(t)}$
as the IRS reflection coefficients at time slot $t$
 and substituting \eqref{superposed} into \eqref{receive}, we obtain
 \vspace{-0.15cm}
\begin{align}\label{receive1}
{\bm y}^{(t)}&= \sum_{k=1}^{K} {\bm X}_k^{(t)} {\bm F} \left( {\bm Q}_k {\bm \theta}^{(t)}+{\bm d}_k\right)+ {\bm v}^{(t)}.
\end{align}

In this paper, the uplink training for the multi-user channel estimation at the AP is based on the pilot signals sent by the users and the time-varying reflection pattern design at the IRS.
Specifically, the uplink training design consists of two parts: the pilot tone allocations $\{\delta_{k,n}^{(t)}\}$ for different users over $\tau$ OFDM pilot symbols and the IRS reflections $\{{\bm \theta}^{(t)}\}$ over different OFDM pilot symbols, both of which need to be carefully designed to minimize the channel estimation error for all the users.
In the following two sections, we present two efficient channel estimation schemes with optimized training designs for different channel setups, respectively, and derive the fundamental limits of these schemes on 
the minimum training overhead 
and the maximum number of supportable users in the IRS-assisted multi-user OFDMA system.
\section{Simultaneous-User Channel Estimation and Training Design for Arbitrary Channels }\label{Concurrent-User}
In this section, we first propose a general channel estimation scheme for the IRS-assisted multi-user OFDMA system under arbitrary channels, where the CSI of all users is estimated in parallel simultaneously at the AP, thus referred to as the simultaneous-user channel estimation (SiUCE) scheme.
For this scheme, the minimum training time, the maximum number of supportable users, and the corresponding optimal joint training design of pilot tone allocations and IRS reflection pattern to minimize the channel estimation error are derived accordingly.
\subsection{Channel Estimation and Maximum Number of Supportable Users}\label{Concurrent_est}
Without loss of generality, we assume that the pilot tones assigned to each user are identical over different time slots, i.e., $\delta_{k,n}^{(t)}=\delta_{k,n},  \forall t \in {\cal T},  \forall n \in {\cal N}, \forall k \in {\cal K}$.
As such, we have ${\cal J}_{k}^{(t)}={\cal J}_{k}$ and ${\bm X}_k^{(t)}={\bm X}_k, \forall t \in {\cal T}, \forall k \in {\cal K}$.
By defining ${\tilde{\bm Q}}_k=\left[{\bm d}_k, {\bm Q}_k \right]$ and $ {\tilde{\bm \theta} }^{(t)}= \begin{bmatrix}1\\{\bm \theta}^{(t)}\end{bmatrix}$, \eqref{superposed} can be written
in a compact form as 
${\bm h}_k^{(t)}={\tilde{\bm Q}}_k {\tilde{\bm \theta} }^{(t)} $ and \eqref{receive1} can be rewritten as 
\begin{align}\label{receive1.5}
{\bm y}^{(t)}&= \sum_{k=1}^{K} {\bm X}_k {\bm F} {\tilde{\bm Q}}_k {\tilde{\bm \theta} }^{(t)} 
+ {\bm v}^{(t)}.
\end{align}
Due to the disjoint pilot tone allocations, the received signal vectors for different users can be decoupled as
\begin{align}\label{receive1.6}
{\bm y}_{k}^{(t)}=&{\bm \Pi}_{{\cal J}_{k}}{\bm y}^{(t)}
\stackrel{(a1)}{=}{\bm \Pi}_{{\cal J}_{k}} \left({\bm X}_k {\bm F} {\tilde{\bm Q}}_k {\tilde{\bm \theta} }^{(t)} + {\bm v}^{(t)}\right)\notag\\
\stackrel{(a2)}{=}&\sqrt{\frac{P}{|{\cal J}_{k}|}} {\bm \Pi}_{{\cal J}_{k} }{\bm F} {\tilde{\bm Q}}_k {\tilde{\bm \theta} }^{(t)}+{\bm \Pi}_{{\cal J}_{k}}{\bm v}^{(t)}\notag\\
=&\sqrt{\frac{P}{|{\cal J}_{k}|}} {\bm F}_{k} {\tilde{\bm Q}}_k {\tilde{\bm \theta} }^{(t)}+{\bm v}_{k}^{(t)}, \quad \forall k \in {\cal K}
\end{align}
where ${\bm \Pi}_{{\cal J}_{k}}$ denotes the sub-carrier selection matrix which consists of the ${|{\cal J}_{k}|}$ rows indexed by ${\cal J}_{k}$ of the identical matrix ${\bm I}_N$,
$(a1)$ holds since ${\bm \Pi}_{{\cal J}_{k}}{\bm X}_{k'}={\bm 0}_{|{\cal J}_{k}|\times N}$ for $k'\ne k$,
$(a2)$ holds since ${\bm \Pi}_{{\cal J}_{k}}{\bm X}_k=\sqrt{\frac{P}{|{\cal J}_{k}|}} {\bm \Pi}_{{\cal J}_{k}}$ according to \eqref{OFDM_sym}, ${\bm F}_{k}={\bm \Pi}_{{\cal J}_{k} }{\bm F}$ denotes the ${|{\cal J}_{k}|} \times L$ matrix consisting of the ${|{\cal J}_{k}|}$ rows indexed by ${\cal J}_{k}$ of ${\bm F}$, and
${\bm v}_{k}^{(t)}={\bm \Pi}_{{\cal J}_{k}}{\bm v}^{(t)}$ is the corresponding AWGN vector on the tones of ${\cal J}_{k}$ at each time slot $t$.

By stacking the received signal vectors $\{{\bm y}_{k}^{(t)}\}$ over time slots ${\cal T}$ into ${\bm Y}_{k}=[{\bm y}_{k}^{(1)},{\bm y}_{k}^{(2)},\ldots,{\bm y}_{k}^{(\tau)}]$, we obtain
\begin{align}\label{Y_mat}
{\bm Y}_{k}=\sqrt{\frac{P}{|{\cal J}_{k}|}} {\bm F}_{k} {\tilde{\bm Q}}_k {\bm \Xi}+{\bm V}_{k}, \qquad  \forall k \in {\cal K}
\end{align}
\rev{where 
${\bm \Xi} \triangleq [{\tilde{\bm \theta} }^{(1)},{\tilde{\bm \theta} }^{(2)},\ldots,{\tilde{\bm \theta} }^{(\tau)}]$ denotes the IRS reflection pattern matrix that collects all reflection coefficients $\{{\tilde{\bm \theta} }^{(t)}\}$ over time slots ${\cal T}$} and ${\bm V}_{k}=[{\bm v}_{k}^{(1)},{\bm v}_{k}^{(2)},\ldots,{\bm v}_{k}^{(\tau)}]$ denotes the corresponding AWGN matrix.
Let ${\bm F}_{k}^\dagger=\left({\bm F}_{k}^H {\bm F}_{k}\right)^{-1}{\bm F}_{k}^H$ and
${\bm \Xi}^\dagger={\bm \Xi}^H\left({\bm \Xi} {\bm \Xi}^H\right)^{-1}$ denote
the left pseudo-inverse of ${\bm F}_{k}$ and the right pseudo-inverse of ${\bm \Xi}$, respectively.
By left- and right-multiplying ${\bm Y}_{k}$ in \eqref{Y_mat} by $\sqrt{\frac{|{\cal J}_{k}|}{P}} {\bm F}_{k}^\dagger$ and ${\bm \Xi}^\dagger$, respectively, we obtain the least-square (LS) estimates of
${\bm d}_k$ and ${\bm Q}_k$ as
\begin{align}\label{LS_est}
\left[{\hat{\bm d}}_k, {\hat{\bm Q}}_k \right]&={\hat{\tilde{\bm Q}}}_k=\sqrt{\frac{|{\cal J}_{k}|}{P}} {\bm F}_{k}^\dagger {\bm Y}_{k} {\bm \Xi}^\dagger\notag\\
&={\tilde{\bm Q}}_k+ \sqrt{\frac{|{\cal J}_{k}|}{P}} {\bm F}_{k}^\dagger {\bm V}_{k}{\bm \Xi}^\dagger, \qquad  \forall k \in {\cal K}
\end{align}
where ${\hat{\bm d}}_k$, ${\hat{\bm Q}}_k$, and ${\hat{\tilde{\bm Q}}}_k$ denote the estimates of
${\bm d}_k$, ${\bm Q}_k$, and ${\tilde{\bm Q}}_k$, respectively.
Note that for the channel estimation based on \eqref{LS_est},
the left pseudo-inverse of ${\bm F}_{k}$ exists if and only if ${\bm F}_{k}$ is of full column rank, which requires 
\begin{align}\label{pilot_user}
{|{\cal J}_{k}|}\ge L, \qquad \forall k\in {\cal K}
\end{align}
and the right pseudo-inverse of ${\bm \Xi}$ exists if and only if ${\bm \Xi}$ is of full row rank, which requires 
\begin{align}\label{tau}
\tau \ge M+1.
\end{align}
From the above, we can infer that for the training overhead of each user $k$, the number of assigned sub-carriers ${|{\cal J}_{k}|}$ should be no less than the maximum delay spread $L$, while the number of OFDM pilot symbols $\tau$ should be no less than the total number of channel links including the direct link and the reflecting links associated with the $M$ sub-surfaces. It is worth pointing out that although \eqref{pilot_user} and \eqref{tau} are the necessary but not necessarily sufficient conditions for achieving the full column rank of ${\bm F}_{k}$ and full row rank of ${\bm \Xi}$, respectively, we claim that a full-column-rank ${\bm F}_{k}$ and a full-row-rank ${\bm \Xi}$ always exist when the conditions in \eqref{pilot_user} and \eqref{tau} are satisfied, which will be specified in the next subsection.
In addition, the number of training time slots $\tau$ should satisfy \eqref{tau} for attaining a unique solution to the estimation based on \eqref{LS_est} and thus the minimum training time is $\tau_{\min}=M+1$.
\rev{To minimize the channel training time, we set $\tau =\tau_{\min}=M+1$ in the rest of this paper.}
Furthermore, according to \eqref{pilot_user} and the disjoint pilot tone allocations for all users, the number of supportable users, $K$, should satisfy the following condition:
\begin{align}
KL \stackrel{(b)}{\le} \sum_{k=1}^K  {|{\cal J}_{k}|}\le N
\end{align}
where the equality of $(b)$ holds if and only if $|{\cal J}_{k}|=L, \forall k\in {\cal K}$. Thus, the maximum number of supportable users by the SiUCE scheme, denoted by $K_1$, is given by
\begin{align}\label{maximum_user}
K_1 = \lfloor{ N} /{L}\rfloor  .
\end{align}

\subsection{Optimal Training Design}\label{Concurrent_design}
Note that the required CSI of each user can be recovered from \eqref{LS_est} when ${\bm F}_{k}$  has full column rank and ${\bm \Xi}$ has full row rank.
However, the matrix inversion operation for computing the pseudo-inverses of ${\bm F}_{k}$ and ${\bm \Xi}$ has a cubic time complexity in general and may lead to suboptimal channel estimation due to the potential noise enhancement if either ${\bm F}_{k}$ or ${\bm \Xi}$ is ill-conditioned. For this sake, in this subsection we optimize the joint training design of the pilot tone allocations for all users and IRS time-varying reflection pattern to minimize the channel estimation error as well as reduce the implementation complexity of the proposed SiUCE scheme.

From \eqref{LS_est}, the average mean square error (MSE) of the SiUCE scheme over the $K$ users is given by
\begin{align}\label{MSE0}
\varepsilon&=\frac{1}{KL(M+1)} \sum_{k=1}^K {\mathbb E}\left\{  \left\|\left[{\hat{\bm d}}_k, {\hat{\bm Q}}_k \right]
-\left[{{\bm d}}_k, {{\bm Q}}_k \right]\right\|^{2}_F
\right\}
\notag \\
&=\frac{1}{KL(M+1)} \sum_{k=1}^K {\mathbb E}\left\{ \Big\|  \sqrt{\frac{|{\cal J}_{k}|}{P}} {\bm F}_{k}^\dagger {\bm V}_{k}{\bm \Xi}^\dagger \Big\|^{2}_F\right\}. 
\end{align}
Accounting for the constraints on the training design, the optimization problem for minimizing the MSE in \eqref{MSE0} is formulated
as follows (with constant/irrelevant terms omitted for brevity).
\begin{align}
\text{(P1):}~
& \underset{ \left\{\theta_m^{(t)}\right\},\left\{\delta_{k,n}\right\}   }{\text{min}}
& & \sum_{k=1}^K {\mathbb E}\left\{ \left\|  \sqrt{\frac{|{\cal J}_{k}|}{P}} {\bm F}_{k}^\dagger {\bm V}_{k}{\bm \Xi}^\dagger \right\|^{2}_F
\right\}\label{obj_P0}\\
& \text{~~~~~s.t.} & & \sum_{k=1}^K \delta_{k,n} \le 1, \forall n \in {\cal N} \label{con2_P0}\\
& & & \delta_{k,n} \in\{0,1\},  \forall n \in {\cal N}, \forall k \in {\cal K}\label{con3_P0}\\
& & & |\theta_m^{(t)}|=1, \forall t \in {\cal T}, \forall  m\in {\cal M}\label{con4_P0}.
\end{align}
It can be verified that problem (P1) is a non-convex optimization problem.
Specifically, the binary constraint in \eqref{con3_P0} and the unit-modulus constraint in \eqref{con4_P0} are non-convex. Moreover, the objective function in \eqref{obj_P0} is non-convex over $\left\{\theta_m^{(t)}\right\}$ and $\left\{\delta_{k,n}\right\}$ via ${\bm \Xi}$ and ${\bm F}_{k}$.
Although the non-convex optimization problem is generally difficult to solve, we obtain the optimal solution to problem (P1) in the following proposition.


\indent\emph{Proposition 1}: 
The optimal solution to problem (P1) for minimizing the MSE of the SiUCE scheme should satisfy:
\begin{itemize}
	\item The optimal IRS reflection pattern ${\bm \Xi}$ is an orthogonal matrix with each entry satisfying the unit-modulus constraint, i.e., ${\bm \Xi}  {\bm \Xi}^H =(M+1){\bm I}_{M+1}$;
	\item The optimal pilot tones allocated to each user $k$ are equispaced over ${|{\cal J}_{k}|}$ sub-carriers with ${|{\cal J}_{k}|}\ge L$ and ${\cal J}_{k} \bigcap {\cal J}_{k'}=\emptyset $ for $k\ne k'$, for which it satisfies ${\bm F}_{k}^H {\bm F}_{k}=\frac{|{\cal J}_{k}|}{N} {\bm I}_{L} , \forall k \in {\cal K}$.
\end{itemize}
Moreover, the minimum MSE is given by
\begin{align}\label{MSE_min}
\varepsilon_{\min}=\frac{\sigma^2 N}{P(M+1)}.
\end{align}
\begin{IEEEproof}
	 Please refer to the Appendix.
\end{IEEEproof}


According to Proposition 1, one optimal training design for the SiUCE scheme is given as follows:
use the $(M+1)\times (M+1)$ DFT matrix as the reflection pattern ${\bm \Xi}$ with each IRS reflection coefficient given by
\begin{align}\label{design_IRS}
\theta_m^{(t)}=e^{-j\frac{2\pi m (t-1) }{M+1}}, \quad \forall m\in {\cal M}, \forall t \in {\cal T}
\end{align}
 and the equispaced pilot tones allocated to each user are indexed by
 \begin{align}\label{design_pilot}
 {\cal J}_{k}=\left\{n \Big|n \mod{\frac{N}{L_p}}=k-1, n \in {\cal N}\right\}, \quad \forall k \in {\cal K}
 \end{align}
where $L_p$ denotes the number of pilot tones allocated to each user at each time slot, which is set to be identical for all the users for fairness (i.e., $|{\cal J}_{k}|=L_p, \forall k \in {\cal K}$) and satisfies $L \le L_p\le \frac{N}{K} $, and the spacing of adjacent pilot tones of each user is $\frac{N}{L_p}$. 
Moreover, given Proposition 1,
we can readily obtain that ${\bm \Xi}^\dagger=\frac{1}{M+1}{\bm \Xi}^H$ and 
${\bm F}_{k}^\dagger=\frac{N}{|{\cal J}_{k}|}{\bm F}_{k}^H, \forall k \in {\cal K}$, both of which dispense with
the matrix inversion operation for reducing the implementation complexity.

\begin{figure}[!t]
	\centering
	\includegraphics[width=2.3in]{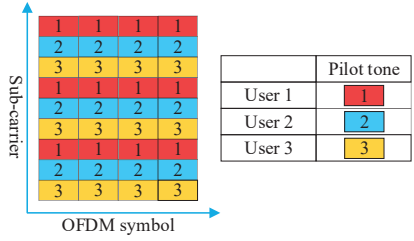}
	\setlength{\abovecaptionskip}{-6pt}
	\caption{Illustration of the equispaced pilot tone allocation design for the SiUCE scheme.}
	\label{peer_pilot}
	\vspace{-0.3cm}
\end{figure}

Last, we give an illustrative example of the proposed equispaced pilot tone allocation design for the SiUCE scheme
in Fig.~\ref{peer_pilot}, with $N=9$, $M=3$, and $L_p=L=3$.
It can be observed that given the minimum channel training time $\tau_{\min}=M+1=4$, the maximum number of supportable users by the SiUCE scheme is $K_1=\lfloor{ N} /{L}\rfloor =3$ in this example.

\section{Sequential-User Channel Estimation and Training Design for LoS Dominant User-IRS Channels}\label{Reference-User}


In this section, we consider a typical scenario where the user-IRS channels are LoS dominant. For this scenario, by exploiting the same (common) IRS-AP channel ${\bar {\bm G}}$ shared by all the users, we propose a customized channel estimation scheme that is capable of supporting more users for channel estimation than the SiUCE scheme proposed in Section \ref{Concurrent-User} which is applicable to arbitrary channels. Specifically, our proposed new channel estimation scheme first estimates the CSI of one arbitrarily selected user, denoted as the reference user, and then recovers the CSI of the remaining non-reference users based on the reference user's CSI, thus referred to as the sequential-user channel estimation (SeUCE) scheme.
For this scheme, the minimum training overhead of each user,
 the maximum number of supportable users, and the corresponding training design for minimizing the channel estimation error are derived as well.
 
For the SeUCE scheme, we consider the case of $L_2=1$ (i.e., all the user-IRS channels are LoS paths) or simply estimate the strongest/dominant time-domain LoS path as an approximation of each user-IRS link for the case of $L_2>1$ (i.e., by ignoring all the non-LoS (NLoS) paths and treating them as noise) to reduce the estimation complexity.
This is usually valid in practice since the distance between each IRS-served user and the IRS is typically short and thus the corresponding channel is dominated by the strong LoS component, while the other NLoS components are much weaker and thus negligible (say, the Rician fading channel with a very high Rician factor).
By slight abuse of notation, we define $L_r= L_1$ and $L=\max\{L_1, L_d \}$, which may be different from those defined in Section \ref{Concurrent-User} due to different channel setups.


Without loss of generality, we assume that the first row of ${\bar {\bm U}}_k$ corresponds to the dominant LoS component of the user-IRS channel for each user $k$ and denote it by ${\bm u}_k^T\triangleq [u_{k,1},u_{k,2},\ldots,u_{k,M}]\in {\mathbb{C}^{1\times M}}$.
As such, the cascaded user-IRS-AP channel matrix (without the effect of phase shifts) can be simplified as (as compared with that given in \eqref{conv_ch}) 
\begin{align}\label{reflect_ch}
{\bm Q}_k &={\bm G} ~\text{diag} \left({\bm u}_k\right)
\end{align}
where ${\bm Q}_k \in {\mathbb{C}^{L\times M }} $, and ${\bm G}$ denotes the zero-padded IRS-AP channel with zero padding length of $L-L_1$ on each column of ${\bar {\bm G}}$.
Then it can be observed that, if given the cascaded user-IRS-AP channel matrix of any user (say, ${\bm Q}_1 ={\bm G} ~\text{diag} \left({\bm u}_1\right)$ of user~$1$),
we can re-express \eqref{reflect_ch} as
\begin{align}\label{reflect_ch2}
{\bm Q}_k &={\bm G} ~\text{diag} \left({\bm u}_k\right)={\bm G} ~\text{diag} \left({\bm u}_1\right)~\left(\text{diag}\left({\bm u}_1\right)\right)^{-1}~ \text{diag} \left({\bm u}_k \right)\notag\\
&= {\bm G} ~\text{diag} \left({\bm u}_1\right)~\text{diag}\left({\bm a}_k\right)={\bm Q}_1 ~\text{diag}\left({\bm a}_k\right)
\end{align}
where $\text{diag}\left({\bm a}_k\right)=\left(\text{diag}\left({\bm u}_1\right)\right)^{-1} \text{diag}\left({\bm u}_k\right)$ is the diagonal user-IRS channel matrix normalized by ${\bm u}_1$,
and we have ${\bm a}_k\in {\mathbb{C}^{M\times 1}}$ and ${\bm a}_1={\bm 1}_{M\times 1}$.
This key observation indicates that given the cascaded user-IRS-AP channel matrix of an arbitrary user, other users' cascaded reflecting CSI can be recovered with the normalized user-IRS channel ${\bm a}_k$, which has a much lower dimension than ${\bm Q}_k$.
As such, without loss of generality,
by taking user $1$ as the reference user and substituting ${\bm Q}_k$ of \eqref{reflect_ch2} into \eqref{superposed}, the superimposed CIR from user $k$ to the AP in the time domain can be rewritten as
\begin{align}\label{superposed3}
\hspace{-0.15cm}{\bm h}_k&={\bm Q}_k {\bm \theta}+{\bm d}_k={\bm Q}_1  \text{diag}\left({\bm a}_k\right) {\bm \theta}+{\bm d}_k={\bm Q}_1 {\bm \Theta} {\bm a}_k+{\bm d}_k\hspace{-0.15cm}
\end{align}
where ${\bm \Theta}=\text{diag} \left( {\bm \theta} \right)$ represents the diagonal reflection matrix of the IRS and ${\bm d}_k\in {\mathbb{C}^{L\times 1}}$ is the zero-padded user-AP direct channel of ${\bar {\bm d}}_k$ with the zero padding length of $L- L_d$.
According to \eqref{superposed3}, it is sufficient to acquire the channel knowledge of the reference user's cascaded reflecting channel ${\bm Q}_1$, the normalized user-IRS channels $\{{\bm a}_k\}_{k=2}^K$, and the direct channels $\{{\bm d}_k\}_{k=1}^K$ for the $K$ users, which inspires us to propose the SeUCE scheme.
It is worth pointing out that for the typical scenario where $L_d\ge L_r$ and thus $L=\max\{L_r, L_d \}=L_d$ is identical for the two proposed channel estimation schemes, namely, SiUCE and SeUCE, 
the number of channel coefficients to be estimated in the SeUCE scheme is significantly reduced to $LM+(K-1)M+KL$ by exploiting the common IRS-AP channel, as compared to the  SiUCE scheme that requires estimating $\{{\bm Q}_k\}_{k=1}^K$ and $\{{\bm d}_k\}_{k=1}^K$ with $(M+1)KL$ coefficients in total, due to the fact that $L+K\ll LK$ in practical OFDMA systems.
After substituting \eqref{superposed3} into \eqref{receive}, the received signal is rewritten as
\begin{align}\label{receive_ref}
{\bm y}^{(t)}&= \sum_{k=1}^{K} {\bm X}_k^{(t)} {\bm F} \left( {\bm Q}_1 {\bm \Theta}^{(t)} {\bm a}_k+{\bm d}_k \right)
+ {\bm v}^{(t)}.
\end{align}

Based on the above discussions (especially the property revealed in \eqref{reflect_ch2}),
the main procedures of the proposed SeUCE scheme are described as follows and will be further elaborated in the subsequent subsections. 
\begin{enumerate}
	\item With the received pilot signals assigned to user 1 (the reference user), 
	we estimate the CSI of ${\bm Q}_1$ and ${\bm d}_1$ for the reference user; 
	\item  With the received pilot signals assigned to users $2$ to $K$ (the remaining non-reference users), we estimate the CSI of $\{{\bm a}_k\}_{k=2}^K$ and $\{{\bm d}_k\}_{k=2}^K$ for the remaining non-reference users and recover each ${\bm Q}_k$ from the estimated ${\bm Q}_1$ and ${\bm a}_k$
	according to \eqref{reflect_ch2}.
\end{enumerate}
For the SeUCE scheme, we further derive 
the minimum training overhead of each user and
the maximum number of supportable users, as well as optimize the corresponding training design for minimizing the channel estimation error of all users. Note that as the channel estimation for the non-reference users is coupled with that for the reference user, the optimal joint training design for all the users is highly challenging in general. To tackle this challenge, we propose a suboptimal training design by decoupling the joint design problem into the following two sub-problems, with details given in the subsequent subsections as well.
\begin{enumerate}
	\item Given the number of pilot tones allocated to user~1 (the reference user), we optimize the pilot tone allocation for the reference user and the IRS reflection pattern ${\bm \Xi}$; 
	\item Given the optimized IRS reflection pattern ${\bm \Xi}$ and the remaining pilot tones (not occupied by the reference user), we optimize the pilot tone allocations jointly for the remaining $K-1$ non-reference users.
\end{enumerate}
\subsection{Channel Estimation and Optimal Training Design for Reference User}\label{ref_design}
\subsubsection{Channel Estimation}

Let ${\cal J}_{1}$ denote the index set of the pilot tones allocated to user 1 (the reference user), which is assumed to be identical over different time slots. Thus, we have ${\cal J}_{1}^{(t)}={\cal J}_{1}$ and ${\bm X}_1^{(t)}={\bm X}_1,  \forall t \in {\cal T}$.  
Similar to the case of $k=1$ in Section \ref{Concurrent_est},
the received signal of the reference user (by collecting the pilot tones of ${\cal J}_{1}$) is expressed as
\begin{align}
{\bm y}^{(t)}_1=&{\bm \Pi}_{{\cal J}_{1}}{\bm y}^{(t)}
\stackrel{(c1)}{=} {\bm \Pi}_{{\cal J}_{1}}{\bm X}_1 {\bm F} \left( {\bm Q}_1 {\bm \theta}^{(t)} +{\bm d}_1 \right)
+ {\bm \Pi}_{{\cal J}_{1}}{\bm v}^{(t)}\notag\\
\stackrel{(c2)}{=}&\sqrt{\frac{P}{|{\cal J}_{1}|}} {\bm F}_1 {\tilde{\bm Q}}_1 {\tilde{\bm \theta} }^{(t)} +{\bm v}_1^{(t)}
\end{align}
where 
$(c1)$ holds since ${\bm \Pi}_{{\cal J}_{1}}{\bm X}_{k}^{(t)}={\bm 0}_{|{\cal J}_{1}|\times N}$ for $k\ne 1$ due to the disjoint pilot tone allocations and ${\bm \theta}^{(t)}={\bm \Theta}^{(t)} {\bm 1}_{M\times 1}={\bm \Theta}^{(t)} {\bm a}_1$, and
$(c2)$ holds since ${\bm \Pi}_{{\cal J}_{1}}{\bm X}_{1}=\sqrt{\frac{P}{|{\cal J}_{1}|}} {\bm \Pi}_{{\cal J}_{1}}$ and ${\bm F}_{1}={\bm \Pi}_{{\cal J}_{1} }{\bm F}$.
By stacking the received signal vectors $\{{\bm y}_{1}^{(t)}\}$ over $M+1$ time slots into ${\bm Y}_{1}=\left[{\bm y}_{1}^{(1)},{\bm y}_{1}^{(2)},\ldots,{\bm y}_{1}^{(M+1)}\right]$, we obtain
\vspace{-0.15cm}
\begin{align}\label{Y_matR}
{\bm Y}_{1}=\sqrt{\frac{P}{|{\cal J}_{1}|}} {\bm F}_{1} {\tilde{\bm Q}}_{1} {\bm \Xi}+{\bm V}_{1} .
\end{align}
Then, left- and right-multiplying ${\bm Y}_{1}$ in \eqref{Y_matR} by $\sqrt{\frac{|{\cal J}_{1}|}{P}} {\bm F}_{1}^\dagger$ and ${\bm \Xi}^{-1}$, respectively, we get the LS estimates of
${\bm d}_{1}$ and ${\bm Q}_{1}$ as follows.
\begin{align}\label{LS_estR}
\left[{\hat{\bm d}}_{1}, {\hat{\bm Q}}_{1} \right]&={\hat{\tilde{\bm Q}}}_{1}=\sqrt{\frac{|{\cal J}_{1}|}{P}} {\bm F}_{1}^\dagger {\bm Y}_{1} {\bm \Xi}^{-1}\notag\\
&={\tilde{\bm Q}}_{1}+ \sqrt{\frac{|{\cal J}_{1}|}{P}} {\bm F}_{1}^\dagger {\bm V}_{1}{\bm \Xi}^{-1}
\end{align}
where ${\hat{\bm d}}_{1}$, ${\hat{\bm Q}}_{1}$, and ${\hat{\tilde{\bm Q}}}_{1}$ denote the estimates of
${\bm d}_{1}$, ${\bm Q}_{1}$, and ${\tilde{\bm Q}}_{1}$ for the reference user, respectively, and
${\bm F}_{1}^\dagger=\left({\bm F}_{1}^H {\bm F}_{1}\right)^{-1}{\bm F}_{1}^H$ is
the left pseudo-inverse of ${\bm F}_{1}$.
Note that for the channel estimation based on \eqref{LS_estR},
the left pseudo-inverse of ${\bm F}_{1}$ exists if and only if ${\bm F}_{1}$ is of full column rank, which requires ${|{\cal J}_{1}|}\ge L$ for the training overhead of the reference user.
\subsubsection{Training Design}
Following a similar procedure for optimizing the training design in Section \ref{Concurrent_design} with $k=1$, we can readily conclude that
the minimum MSE of the channel estimation in \eqref{LS_estR} for the reference user can be achieved
when the IRS reflection pattern ${\bm \Xi}$ is an orthogonal matrix with each entry satisfying the unit-modulus constraint, i.e., ${\bm \Xi}  {\bm \Xi}^H =(M+1){\bm I}_{M+1}$, and the pilot tones assigned to the reference user are equispaced with ${|{\cal J}_{1}|}\ge L$, for which it satisfies ${\bm F}_{1}^H {\bm F}_{1}=\frac{|{\cal J}_{1}|}{N} {\bm I}_{L} $.
Moreover, 
one optimal training design can be obtained according to \eqref{design_IRS} and \eqref{design_pilot} with $k=1$, and the corresponding minimum MSE is given by $\varepsilon_{\rm{ref}}=\frac{\sigma^2 N}{P(M+1)}$.
\subsection{Channel Estimation for Non-reference Users and Maximum Number of Supportable Users}
After acquiring the CSI of ${\bm Q}_1$ from \eqref{LS_estR}, we then estimate 
the normalized user-IRS channel ${\bm a}_k$ to recover
the cascaded reflecting channel ${\bm Q}_k$ for each non-reference user according to \eqref{reflect_ch2}.
As the pilot tones of ${\cal J}_{1}$ have been occupied by the reference users, we set
$\delta_{k,n}^{(t)}=0, \forall t \in {\cal T} , \forall n\in {\cal J}_{1}, \forall k \in {\bar{\cal K}} \triangleq {\cal K} \setminus\{1\}$ for the remaining non-reference users.
Due to the disjoint pilot tone allocations,
the received signal vector for each of the remaining $K-1$ non-reference users can be expressed as
\begin{align}
{{\bm z}}_{k}^{(t)}
=&{\bm \Pi}_{{\cal J}_{k}^{(t)}}  {\bm y}^{(t)}=  {\bm \Pi}_{{\cal J}_{k}^{(t)}} {{\bm X}}_k^{(t)} {{\bm F}} \left( {\bm Q}_1 {\bm \Theta}^{(t)} {\bm a}_k+{\bm d}_k \right)
+ {\bm \Pi}_{{\cal J}_{k}^{(t)}}  {\bm v}^{(t)}\notag\\
\stackrel{(d)}{=}& \sqrt{\frac{P}{|{\cal J}_{k}^{(t)}|}} 
 {{\bm F}}_{k}^{(t)} \left( {\bm Q}_1 {\bm \Theta}^{(t)} {\bm a}_k+{\bm d}_k \right) + {\bm v}_{k}^{(t)}\label{receive_other0}\\
 =&   {{\bm C}}_k^{(t)} {\bm \lambda}_k+ {\bm v}_{k}^{(t)}
 , \qquad  \forall k \in {\bar{\cal K}}\label{receive_other}
\end{align}
where ${\bm \Pi}_{{\cal J}_{k}^{(t)}}$ denotes the sub-carrier selection matrix which consists of the ${|{\cal J}_{k}^{(t)}|}$ rows indexed by ${\cal J}_{k}^{(t)}$ of the identical matrix ${\bm I}_N$,
$(d)$ holds since ${\bm \Pi}_{{\cal J}_{k}^{(t)}}{\bm X}_{k}^{(t)}=\sqrt{\frac{P}{|{\cal J}_{k}^{(t)}|}} {\bm \Pi}_{{\cal J}_{k}^{(t)}}$ and ${{\bm F}}_{k}^{(t)}={\bm \Pi}_{{\cal J}_{k}^{(t)}} {{\bm F}}$,
${\bm \lambda}_k\triangleq\begin{bmatrix} {\bm a}_k\\{\bm d}_k\end{bmatrix}$, 
${{\bm C}}_k^{(t)}\triangleq \sqrt{\frac{P}{|{\cal J}_{k}^{(t)}|}} {{\bm F}}_{k}^{(t)} \left[ {\bm Q}_1 {\bm \Theta}^{(t)},~ {\bm I}_L\right]$, and ${\bm v}_{k}^{(t)}={\bm \Pi}_{{\cal J}_{k}^{(t)}} {\bm v}^{(t)}$
is the corresponding AWGN vector on the pilot tones of ${\cal J}_{k}^{(t)}$ at each time slot $t$.

By collecting the received signal vectors $\{{{\bm z}}_{k}^{(t)}\}$ of non-reference user $k$ over $M+1$ time slots into ${{\bm z}}_{k}=\left[({{\bm z}}_{k}^{(1)})^T,\ldots,({{\bm z}}_{k}^{(M+1)})^T\right]^T$, we obtain
\begin{align}\label{receive_z}
{{\bm z}}_{k}&=   {{\bm C}}_{k} {\bm \lambda}_k +{\bm v}_{k}, \qquad  \forall k \in {\bar{\cal K}}
\end{align}
where ${\bm v}=\left[({\bm v}^{(1)})^T,\ldots,({\bm v}^{(M+1)})^T\right]^T$ and 
\begin{align}\label{es_mat2}
\hspace{-0.3cm}{{\bm C}}_k\hspace{-0.11cm}
=\hspace{-0.2cm}\begin{bmatrix}
{{\bm C}}_k^{(1)}\\
\vdots \\
{{\bm C}}_k^{(M+1)}
\end{bmatrix}
\hspace{-0.2cm}=\hspace{-0.2cm}\begin{bmatrix}
\sqrt{\frac{P}{|{\cal J}_{k}^{(1)}|}}{{\bm F}}_{k}^{(1)} \left[ {\bm Q}_1 {\bm \Theta}^{(1)},~ {\bm I}_L\right]\\
\vdots \\
\hspace{-0.1cm}\sqrt{\frac{P}{|{\cal J}_{k}^{(M+1)}|}}{{\bm F}}_{k}^{(M+1)} \hspace{-0.1cm}\left[ {\bm Q}_1 {\bm \Theta}^{(M+1)},~ {\bm I}_L\right]\hspace{-0.1cm}
\end{bmatrix}\hspace{-0.1cm}.\hspace{-0.1cm}
\end{align}
Let ${{\bm C}}_k^\dagger=\left({{\bm C}}_k^H {{\bm C}}_k\right)^{-1}{{\bm C}}_k^H$
denote
the left pseudo-inverse of ${{\bm C}}_k$.
Then, left-multiplying ${{\bm z}}_{k}$ in \eqref{receive_z} by ${{\bm C}}_k^\dagger$, we obtain the LS estimates of
${\bm a}_k$ and ${\bm d}_k$ as
\begin{align}\label{LS_estZ}
\begin{bmatrix} {\hat{\bm a}}_k\\ {\hat{\bm d}}_k\end{bmatrix}&={\hat{\bm \lambda}}_k={{\bm C}}_k^\dagger {{\bm z}}_{k}={\bm \lambda}_k+ {{\bm C}}_k^\dagger{\bm v}_{k}, \qquad  \forall k \in {\bar{\cal K}}
\end{align}
where ${\hat{\bm a}}_k$ ${\hat{\bm d}}_k$, and ${\hat{\bm \lambda}}_k$ denote the estimates of
${\bm a}_k$, ${\bm d}_k$, and ${\bm \lambda}_k$, respectively.
Note that for the channel estimation based on \eqref{LS_estZ},
the left pseudo-inverse of ${{\bm C}}_k$ exists if and only if ${{\bm C}}_k$ is of full column rank, which requires 
\begin{align}\label{inv_B}
\zeta_{k} \triangleq \sum_{t=1}^{M+1}\left|{\cal J}_{k}^{(t)}\right| \ge M+L, \qquad  \forall k \in {\bar{\cal K}}
\end{align}
where $\zeta_{k}$ denotes the total number of pilot tones assigned to non-reference user $k$ (i.e., training overhead), which should be no less than $M+L$.
Similarly, although \eqref{inv_B} is a necessary but generally not sufficient condition for achieving the full column rank of ${{\bm C}}_k$, a full-column-rank matrix ${{\bm C}}_k$ exists when the condition in \eqref{inv_B} is met, which will be specified in the next subsection.
Moreover, due to the disjoint pilot tone allocations for the non-reference users, 
we have
\begin{align}\label{disjoint_con}
\sum_{k=2}^{K}\left|{\cal J}_{k}^{(t)}\right| \le N-{|{\cal J}_{1}|},\qquad  \forall t \in {\cal T}.
\end{align}
By combining \eqref{inv_B} and \eqref{disjoint_con}, we arrive at
the following condition on the number of supportable users by the SeUCE scheme (recall that ${|{\cal J}_{1}|}\ge L$ pilot tones at each time slot are required for the reference user):
\begin{align}
&(K-1)(M+L) \stackrel{(e1)}{\le} \sum_{t=1}^{M+1}\sum_{k=2}^{K}\left|{\cal J}_{k}^{(t)}\right| \notag\\
\le&
(M+1)( N-{|{\cal J}_{1}|})
\stackrel{(e2)}{\le} (M+1)( N-L)
\end{align}
where the equality of $(e1)$ holds if and only if $\zeta_{k}=\sum_{t=1}^{M+1}\left|{\cal J}_{k}^{(t)}\right| = M+L, \forall k\in {\bar{\cal K}}$, and the equality of $(e2)$ holds if and only if $|{\cal J}_{1}|=L$.
As a result, the maximum number of supportable users by the SeUCE scheme, denoted by $K_2$, is given by
\begin{align}\label{maximum_user_ref}
K_2 = \left\lfloor \frac{(M+1) (N-L)}{M+L} \right\rfloor +1 .
\end{align}
By comparing \eqref{maximum_user} and \eqref{maximum_user_ref} and assuming that the variables of the floor function $\lfloor \cdot \rfloor$ are integers in both of them, we have
\begin{align}\label{gap}
K_2-K_1&=\frac{(M+1) (N-L)}{M+L}+1- \frac{ N} {L} \notag\\
&=\frac{M(N-L)(L-1)}{(M+L)L} \stackrel{(f)}{\ge} 0
\end{align}
where the equality of $(f)$ holds if and only if $L=1$ or $L=N$, which implies that the maximum number of supportable users by the SeUCE scheme is always no less than that by the SiUCE scheme.
Note that $L=1$ corresponds to the case in which all the involved user-AP, user-IRS, and IRS-AP links are frequency-flat fading channels with one (equivalent) single path (e.g., LoS channels), while $L=N$ is impossible for practical OFDMA systems.
Moreover, \eqref{gap} provides direct insight into the effects of different parameters ($N$, $M$, and $L$) on the relationship between the two channel estimation schemes in terms of maximum number of supportable users.

\rev{\indent\emph{Remark 1}: Note that for the general case with non-negligible multi-path delay spread in the user-IRS link, how to exploit the common IRS-AP channel for all users to fully recover the (exact) channels of the non-reference users based on the estimated CSI of the reference user is highly challenging and still remains open, due to the convolution of the user-IRS and (common) IRS-AP channels, as given in \eqref{conv_ch}. Nevertheless, the proposed SeUCE scheme is still applicable by only estimating the dominant path of each user-IRS link, while the effect of multi-path interference in the user-IRS link on the MSE performance will be evaluated by simulations in~Section \ref{Sim}.}

\subsection{Pilot Tone Allocation for Non-reference Users}
In this subsection, we aim to minimize the average MSE for the remaining $K-1$ non-reference users by jointly
optimizing the corresponding pilot tone allocations.
From \eqref{LS_estZ}, the average MSE of the LS channel estimation over the $K-1$ non-reference users is derived as
	\begin{align}\label{MSE2}
\hspace{-0.3cm}\varepsilon_{\rm non}&\hspace{-0.1cm}=\hspace{-0.1cm}\frac{1}{(M+L)(K-1)} \sum_{k=2}^{K} {\mathbb E}\left\{  \left\|\hat{\bm \lambda}_k
-{\bm \lambda}_k\right\|^{2}
\right\}\notag\\
&\hspace{-0.1cm}=\hspace{-0.1cm}\frac{1}{(M+L)(K-1)} \sum_{k=2}^{K} {\mathbb E}\left\{  \left\|{{\bm C}}_k^\dagger{\bm v}_{k}\right\|^{2}
\right\}\notag \\
&\hspace{-0.1cm}=\hspace{-0.1cm}\frac{1}{(M+L)(K-1)} \sum_{k=2}^{K} \text{tr}\hspace{-0.05cm}\left\{\hspace{-0.1cm} {{\bm C}}_k^\dagger 
{\mathbb E}\left\{ {\bm v}_{k} {\bm v}_{k}^H  \hspace{-0.1cm}\right\}
\left({{\bm C}}_k^\dagger \right)^H	\right\}.
\end{align}
Since ${\mathbb E}\left\{ {\bm v}_{k} {\bm v}_{k}^H  \right\}=\sigma^2 {\bm I}_{\zeta_{k}}$, 
the average MSE in \eqref{MSE2} can be written as
\begin{align}\label{MSE3}
\varepsilon_{\rm non}&=\frac{\sigma^2}{(M+L)(K-1)} \sum_{k=2}^{K} \text{tr}\left\{ {{\bm C}}_k^\dagger
\left( {{\bm C}}_k^\dagger \right)^H	\right\}\notag\\
&=\frac{\sigma^2}{(M+L)(K-1)} \sum_{k=2}^{K} \text{tr}\left\{ \left( {{\bm C}}_k^H{{\bm C}}_k\right)^{-1}	\right\}.
\end{align}
Moreover, according to \eqref{es_mat2}, we have
\begin{align}
&{\bm D}_{k}\triangleq  {{\bm C}}_k^H{{\bm C}}_k=\sum_{t=1}^{M+1} ({{\bm C}}_k^{(t)})^H{{\bm C}}_k^{(t)} \notag\\ 
\hspace{-0.3cm}=&\hspace{-0.15cm}\sum_{t=1}^{M+1}\hspace{-0.15cm}  
{\frac{P}{|{\cal J}_{k}^{(t)}|}}
\begin{bmatrix}  ({\bm \Theta}^{(t)})^H {\bm Q}_1^H \\{\bm I}_L\end{bmatrix}
   ({{\bm F}}_{k}^{(t)})^H {{\bm F}}_{k}^{(t)} \left[ {\bm Q}_1 {\bm \Theta}^{(t)},~ {\bm I}_L\right]\notag\\
   \hspace{-0.3cm}\stackrel{(g)}{=}&\hspace{-0.15cm}\sum_{t=1}^{M+1} \hspace{-0.15cm}
   {\frac{P}{|{\cal J}_{k}^{(t)}|}}\hspace{-0.1cm}
   \sum_{n \in {\bar{\cal N}}} \hspace{-0.1cm}\delta_{k,n}^{(t)}\hspace{-0.1cm}
  \begin{bmatrix}  ({\bm \Theta}^{(t)})^H {\bm Q}_1^H {\bar{\bm f}}_n \\{\bar{\bm f}}_n\end{bmatrix}
  \left[ {\bar{\bm f}}_n^H {\bm Q}_1 {\bm \Theta}^{(t)},~ {\bar{\bm f}}_n^H\right] \label{D_channel0}
\end{align}
where ${\bar{\cal N}}\triangleq {\cal N}\setminus {\cal J}_{1}$,
${\bar{\bm f}}_n^H \in {\mathbb{C}}^{1\times L}$ denotes the $n$-th row vector of ${\bm F}$,
and $(g)$ holds since $({{\bm F}}_{k}^{(t)})^H {{\bm F}}_{k}^{(t)} = \sum_{n \in {\bar{\cal N}}} \delta_{k,n}^{(t)} {\bar{\bm f}}_n {\bar{\bm f}}_n^H$.
To guarantee the feasibility of the LS channel estimation based on \eqref{LS_estZ}, each ${\bm D}_{k} \in {\mathbb{C}^{(M+L)\times (M+L)}}$ should be of full rank. However, it is difficult to obtain the explicit constraints on the pilot tone allocations for the non-reference users, i.e., $\{\delta_{k,n}^{(t)}\}$, to guarantee the full rank of ${\bm D}_{k}$, which can be observed from \eqref{D_channel0}. To overcome this difficulty, we first present an important conjecture as follows.

\indent\emph{Conjecture 1}: Assuming that the channel realization ${\bm Q}_1$ is a random matrix,
each $(M+L)\times (M+L) $ matrix ${\bm D}_{k}$ is of full rank with probability $1$
 if the following conditions are satisfied:
\begin{align}
\sum_{n \in {\bar{\cal N}}} \delta_{k,n}^{(t)} &\ge 1, \qquad\qquad  \forall t \in {\cal T},\forall k \in {\bar{\cal K}} \label{req1_D}\\
\left|\bigcup_{t=1}^{M+1} {\cal J}_{k}^{(t)} \right|&\ge L, \qquad\qquad \forall k \in {\bar{\cal K}}\label{req2_D}\\
\sum_{t=1}^{M+1} \sum_{n \in {\bar{\cal N}}} \delta_{k,n}^{(t)} &\ge M+L, \qquad \forall k \in {\bar{\cal K}}\label{req3_D}.
\end{align}
In the above, \eqref{req1_D} is required for estimating the normalized user-IRS channel ${\bm a}_k$ in the absence of interference from the user-AP direct channel ${\bm d}_k$, i.e., ${\bm d}_k={\bm 0}_{L\times 1}$ in \eqref{receive_other0};
\eqref{req2_D} is required for estimating the user-AP direct channel ${\bm d}_k$ in the absence of interference from the user-IRS channel ${\bm u}_k^T$, i.e., ${\bm u}_k^T={\bm 0}_{1\times M}$ and thus ${\bm a}_k={\bm 0}_{M\times 1}$ in \eqref{receive_other0}; and
\eqref{req3_D} is readily derived from \eqref{inv_B} for jointly estimating ${\bm a}_k$ and ${\bm d}_k$ based on \eqref{receive_other}.
In particular, by extensive simulations (more than 10,000 random channel realizations of ${\bm Q}_1$), we observe that ${\bm D}_{k}$ is always of full rank when the pilot tone allocations for the non-reference users meet the conditions given in \eqref{req1_D}-\eqref{req3_D}, which numerically verifies Conjecture~1, while the rigorous proof for it is still unknown based on our best knowledge and thus will be left for our future work.

Conjecture~1 provides the design constraints for the pilot tone allocations of the non-reference users.
On the other hand, since the exact information of ${\bm Q}_1$ in \eqref{D_channel0} is unavailable
prior to designing the pilot tone allocations,
we instead aim to minimize the MSE in \eqref{MSE3} averaged over ${\bm Q}_1$,
which is formulated as follows (with constant/irrelevant terms omitted for brevity).
\begin{align}
\text{(P2):}~
& \underset{\left\{\delta_{k,n}^{(t)}\right\}   } {\text{min}}
& & \sum_{k=2}^{K} \mathbb{E}_{{\bm Q}_1} \left\{\text{tr}\left\{ {\bm D}_{k}^{-1}	\right\}\right\} \label{obj_P2.1}\\
& \text{~~~~s.t.} & &\sum_{k=2}^K \delta_{k,n}^{(t)} \le 1, \forall t \in {\cal T},  \forall n \in {\bar{\cal N}} \label{con2_P2.1}\\
& & & \delta_{k,n}^{(t)} \in\{0,1\}, \forall t \in {\cal T},  \forall n \in {\bar{\cal N}}, \forall k \in {\bar{\cal K}}\label{con3_P2.1}\\
& & & \eqref{req1_D}-\eqref{req3_D}.\notag
\end{align}
It can be verified that problem (P2) is a non-convex combinatorial optimization problem due to the binary constraints. Moreover, due to the lack of the distribution knowledge of ${\bm Q}_1$ and the matrix inversion involved in the objective function, a closed-form expression for the objective function \eqref{obj_P2.1} in problem (P2) is intractable, which makes problem (P2) difficult to solve.


To overcome such difficulty and draw useful insights into the pilot tone allocation design for the non-reference users, we first consider some simple system setups with small $N$ and/or $M$ for the SeUCE scheme,
for which we are able to perform a brute-force search for all possible pilot tone allocations for the non-reference users and retain those allocation patterns that achieve the minimum MSE of \eqref{obj_P2.1}. 
Note that due to the lack of a closed-form expression for \eqref{obj_P2.1},
the expectation of \eqref{obj_P2.1} is calculated based on the Monte-Carlo method.
Then, by learning the structure of the obtained optimal solutions to problem (P2) under these simple system setups, we propose a low-complexity yet efficient pilot tone allocation design for the non-reference users. 
Specifically, for each non-reference user $k$, the allocation of $\zeta_{k}$ pilot tones includes the following two steps:
\begin{enumerate}
	\item Assign ${\tilde L}_{p,k} \triangleq \lfloor\frac{\zeta_{k}-L+1}{M+1}\rfloor$ sub-carriers over $M+1$ time slots to non-reference user $k$, totally ${\tilde L}_{p,k}(M+1)$ pilot tones;
	\item Assign the remaining $\zeta_{k}-{\tilde L}_{p,k}(M+1)$ pilot tones over different unassigned sub-carriers at one time slot to non-reference user $k$.
\end{enumerate}

\begin{figure}[!t]
	\centering
	\includegraphics[width=2.3in]{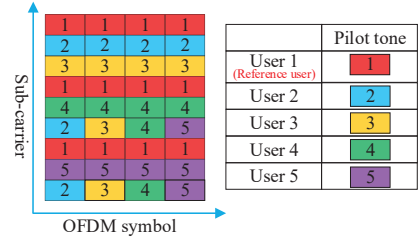}
	\setlength{\abovecaptionskip}{-6pt}
	\caption{Illustration of the proposed pilot tone allocation design for the SeUCE scheme.}
	\label{refer_pilot}
	\vspace{-0.5cm}
\end{figure}

Note that the above design can be applied to a system of arbitrary size (i.e., any values of $N$ and $M$). Next, we give an illustrative example of the proposed pilot tone allocation design for the SeUCE scheme
in Fig. \ref{refer_pilot}, with the same system setup as in Fig.~\ref{peer_pilot}, i.e., $N=9$, $M=3$, and $L=3$.
It can be observed that given the (same) minimum training time $\tau_{\min}=M+1=4$, the maximum number of supportable users by the SeUCE scheme is $ K_2=\lfloor\frac{(M+1) (N-L)}{M+L}\rfloor+1=5$, which is larger than that by the SiUCE scheme (i.e., $K_1=3$) in Section~\ref{Concurrent-User}.

\begin{table*}[!t]
	\begin{center}\caption{Comparison of Two Proposed Channel Estimation Schemes}\label{Table of estimation}
   		\begin{tabular}{|p{23pt}|p{160pt}|p{85pt}|p{40pt}|p{60pt}|}
			\hline
			& Complexity (in terms of average number of complex multiplications)         & Maximum number of supportable users & \multicolumn{2}{p{118pt}|}{Minimum number of pilot tones for each user} \\ \hline
			\hspace{-0.2cm}SiUCE                  & $L(M+1)(L+M+1)\hspace{-0.1cm}\sim\hspace{-0.1cm} {\cal O}((M+1)^2)$          & $K_1 \hspace{-0.1cm}=\hspace{-0.1cm}  \lfloor \frac{ N} {L}\rfloor $                             & \multicolumn{2}{l|}{$(M+1)L$}                            \\ \hline
			\multirow{2}{*}{\begin{tabular}[c]{@{}c@{}}  \\ \hspace{-0.2cm}SeUCE \end{tabular}} & \multirow{2}{*}{\begin{tabular}[c]{@{}c@{}}$\frac{(K-1)\left(2 LM (2M+L+1)+(M+1)^3+7(M+1)^2      \right)}{2K} $
					\\ $+\frac{L(M+1)(L+M+1)}{K}\sim {\cal O}((M+1)^3) $ \end{tabular}} & \multirow{2}{*}{\begin{tabular}[c]{@{}c@{}}  \\ $K_2 \hspace{-0.1cm}=\hspace{-0.1cm} \lfloor\hspace{-0.1cm}\frac{(M+1) (N-L)}{M+L}\hspace{-0.1cm}\rfloor\hspace{-0.1cm}+\hspace{-0.1cm}1 $ \end{tabular} }            & Reference user         & Non-reference user          \\ \cline{4-5} 
			&                    &                                   & $(M+1)L$                 & $M+L$                       \\ \hline
		\end{tabular}
	\end{center}
	\vspace{-0.7cm}
\end{table*}
The comparison between the two proposed channel estimation schemes is summarized in Table~\ref{Table of estimation}. Note that when the number of users $K$ is in the range of $1\le K \le K_1$, we should  adopt the SiUCE for simplicity; while when the number of users $K$ is in the range of $K_1+1\le K \le K_2$, we should adopt the SeUCE for supporting more users at the cost of higher complexity.


\section{Simulation Results}\label{Sim}
In this section, we present simulation results to numerically
validate the effectiveness of our proposed channel estimation schemes as well as their corresponding training designs.
The IRS consists of $M_0=16\times 8=128$ reflecting elements with half-wavelength spacing and is divided into $M=8$ sub-surfaces, each with $\eta= {M}_0/M=16$ elements.
For the purpose of exposition, we consider the uplink training over $\tau_{\min}=M+1=9$ consecutive OFDM symbols and each OFDM symbol consists of $N=16$ sub-carriers appended by a CP of length
$L_{cp}=6$. Moreover, the maximum delay spreads of both the user-AP (direct) channel and the cascaded user-IRS-AP (reflecting) channel are set as $L_r=L_d=4$ and thus $L=\max\{L_r, L_d \}=4$, while the exact settings of $L_1$ and $L_2$ for the IRS-AP and user-IRS channels will be specified later depending
on the scenarios.
Accordingly, the maximum numbers of supportable users by the SiUCE and SeUCE schemes are $K_1=\lfloor \frac{ N} {L}\rfloor=4$ and $K_2=\lfloor\frac{(M+1) (N-L)}{M+L}\rfloor+1=10$, respectively.
The distance-dependent channel path loss is modeled as $\gamma=\gamma_0/ D^\alpha$, where $\gamma_0$ denotes the reference path loss at the reference distance of 1 meter (m),
$D$ denotes the individual link distance, and $\alpha$ denotes the path loss exponent.
The SNR of each user is defined as the ratio between the average power of the received pilot tone and the noise power at the AP, which is given by
\begin{align}
\text{SNR}\hspace{-0.1cm}=\hspace{-0.1cm} {\mathbb E}\hspace{-0.1cm}\left\{\hspace{-0.15cm}\frac{P \left\|{\bm Q}_k {\bm \theta}+{\bm d}_k\right\|^2  }{\sigma^2N} \hspace{-0.15cm}\right\}\hspace{-0.1cm}= \hspace{-0.1cm}\frac{P( M_0\gamma_0^2 D_1^{-\alpha_1}D_2^{-\alpha_2}\hspace{-0.1cm}+\hspace{-0.1cm}\gamma_0D_3^{-\alpha_3} )   }{\sigma^2N}\notag
\end{align}
where $D_1$, $D_2$, and $D_3$ denote the distances of the user-IRS, IRS-AP, and (direct) user-AP links, respectively, $\alpha_1$, $\alpha_2$, and $\alpha_3$ denote the path loss exponents of these links, which are set as $2.2$, $2.4$, and $3.5$, respectively, the path loss at the reference distance $\gamma_0=-30$~dB for each individual link, and the \rev{noise power is set as $\sigma^2=-80$~dBm.}
The distance between the IRS and AP is $50$~m and the users are located on a semi-circle around
the IRS with distance of $1.5$ m, similarly as in \cite{Yang2020IRS}.

For the user-AP and IRS-AP links, the frequency-selective fading channel is modeled by an exponentially decaying power delay profile with a root-mean-square delay spread, where each tap is
generated according to Rayleigh fading and the spread power decaying factor is $2$.
For each user-IRS link modeled by the frequency-selective Rician fading channel (i.e., $L_2>1$), the first tap is set as the LoS component and the remaining taps are NLoS Rayleigh fading  components, with $\kappa$ being the Rician factor that is defined as the ratio of signal power in the dominant LoS component over the total scattered power in NLoS components.
We calculate the normalized MSE over $10,000$ independent fading channel realizations, which is given by
\begin{align}
{\bar\varepsilon}=\frac{1}{KL(M+1)}  \sum_{k=1}^K    {\mathbb E}\left\{\left\| {\hat{\tilde{\bm Q}}}_k-{\tilde{\bm Q}}_k\right\|^{2}_F \Big/{ \left\|{\tilde{\bm Q}}_k\right\|^{2}_F} \right\} .
\end{align}
Note that for the SiUCE scheme,
${\hat{\tilde{\bm Q}}}_k=\left[{\hat{\bm d}}_k, {\hat{\bm Q}}_k \right]$ is obtained according to \eqref{LS_est}, while
for the SeUCE scheme, we obtain ${\hat{\bm Q}}_k={\hat{\bm Q}}_1 ~\text{diag}\left({\hat{\bm a}}_k\right)$ with ${\hat{\bm Q}}_1$ and ${\hat{\bm a}}_k$ given in \eqref{LS_estR} and \eqref{LS_estZ}, respectively, $\forall k \in {\bar{\cal K}}$.

\begin{figure}
	\centering
	\subfigure[Pilot tone allocation benchmark design 1 for the SiUCE.]{
		\begin{minipage}[b]{0.48\textwidth}
			\centering
			\includegraphics[width=2.3in]{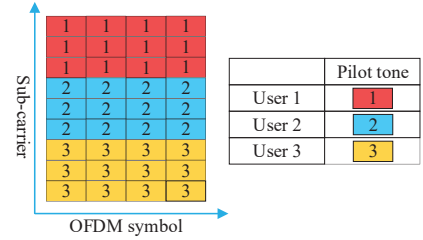}
		\end{minipage}\label{peer_pilot_adj}
	}\\
	\subfigure[Pilot tone allocation benchmark design 2 for the SeUCE.]{
		\begin{minipage}[b]{0.48\textwidth}
			\centering
			\includegraphics[width=2.3in]{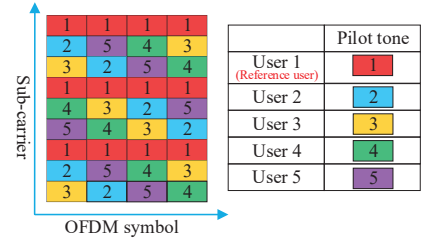}
		\end{minipage}\label{refer_pilot_bench}
	}
	\setlength{\abovecaptionskip}{-3pt}
	\caption{Illustrations of two benchmark designs for pilot tone allocations.} \label{case1}
	\vspace{-0.5cm}
\end{figure}
For the pilot tone allocations, we consider the following two benchmark designs for the proposed SiUCE and SeUCE schemes, respectively. 
\begin{itemize}
	\item {\bf Pilot Tone Allocation Benchmark Design 1 (Adjacent Pilot Tone Allocation):} As shown in Fig.~\ref{peer_pilot_adj}, we consider a heuristic benchmark pilot tone allocation design for the SiUCE scheme, where each user is allocated with $L_p$ adjacent pilot tones indexed by
	${\cal J}_{k}=\left\{(k-1)L_p,(k-1)L_p+1\ldots, kL_p-1\right\}, \forall k \in {\cal K}$ with $L_p$ given in \eqref{design_pilot}.
	\item {\bf Pilot Tone Allocation Benchmark Design 2 (Permutated Pilot Tone Allocation):} As shown in Fig.~\ref{refer_pilot_bench}, we consider another heuristic benchmark pilot tone allocation design for the SeUCE scheme, where the same equispaced pilot tones are allocated to the reference user as that in Section \ref{ref_design}, while the pilot tones assigned to each of the remaining non-reference users are permuted over different sub-carriers and different time slots, which satisfies the conditions in \eqref{req1_D}-\eqref{req3_D} as well.
\end{itemize}
For the IRS reflection pattern over different time slots, besides the proposed DFT-based reflection pattern for the SiUCE and SeUCE schemes, we also consider two benchmark designs as follows. 
\begin{itemize}
	\item {\bf ON/OFF-based Reflection Pattern:} The ON/OFF-based reflection pattern proposed in \cite{yang2019intelligent} is considered for comparison, where the direct channels of all users are estimated first with all the IRS sub-surfaces turned OFF (i.e., $\beta_m=0, \forall m$) in the first time slot, and the reflecting channels are then estimated with one out of $M$ sub-surfaces (say, $i$) turned ON (i.e., $\beta_i=1$ and $\beta_m=0, \forall m\neq i$) sequentially in the remaining time slots. Note that this reflection pattern design is only applicable for the SiUCE scheme.
	\item {\bf Random Reflection Pattern:} The IRS reflection coefficients at each
	time slot are generated with random
	phase shifts (uniformly distributed within $[0, 2\pi)$) and the maximum reflection amplitude (i.e., $\beta_m=1, \forall m$), which are known at the AP for channel estimation. Note that this reflection pattern design is  applicable for both the SiUCE and SeUCE schemes.
\end{itemize}

\begin{figure}
	\centering
	\subfigure[Comparison of different pilot tone allocations.]{
		\begin{minipage}[b]{0.48\textwidth}
			\centering
			\includegraphics[width=2.7in]{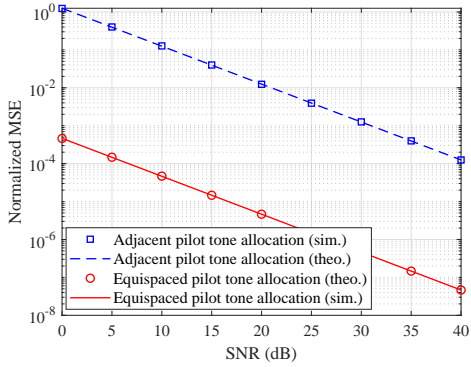}
		\end{minipage}\label{peer_placement}
	}\\
	\subfigure[Comparison of different IRS reflection patterns.]{
		\begin{minipage}[b]{0.48\textwidth}
			\centering
			\includegraphics[width=2.7in]{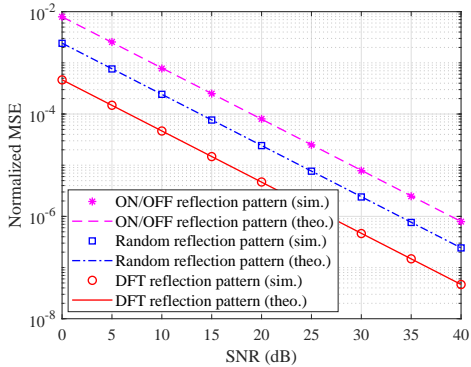}
		\end{minipage}\label{peer_IRSpattern}
	}
	\setlength{\abovecaptionskip}{-3pt}
	\caption{Normalized MSE of the SiUCE scheme versus SNR with $\kappa=4.5$ dB, $L_1=3$, and $L_2=2$.} 
	\vspace{-0.1cm}
\end{figure}

%
%

In Figs.~\ref{peer_placement} and \ref{peer_IRSpattern}, we compare the normalized MSE of different pilot tone allocations and IRS reflection patterns for the SiUCE scheme with $\kappa=4.5$ dB, $L_1=3$, and $L_2=2$. It is observed that the theoretical analysis of MSE given in \eqref{obj_MSE2} is in
agreement with the simulation results.
 Moreover, compared to the benchmark schemes, our proposed equispaced pilot tone allocation and DFT-based reflection pattern jointly achieve the minimum MSE as shown in \eqref{MSE_min}.
Specifically, given the same DFT-based reflection pattern, our proposed equispaced pilot tone allocation design achieves substantial SNR gains over the adjacent pilot tone allocation benchmark due to the ill-conditioned $\{{\bm F}_{k}\}$ in the latter case. On the other hand, given the same equispaced pilot tone allocation, our proposed DFT-based reflection pattern achieves about $12$~dB SNR gain over the ON/OFF-based reflection benchmark  without fully utilizing the large aperture of IRS and $7$~dB SNR gain over the random reflection benchmark due to the noise enhancement after random matrix inversion. 
 Therefore,
the choices of pilot tone allocation and/or IRS reflection pattern have a significant impact
on the MSE performance of the proposed SiUCE scheme.

%

\begin{figure}
	\centering
	\subfigure[Normalized MSE of different pilot tone allocations versus SNR with $L_1=4$ and $L_2=1$.]{
		\begin{minipage}[b]{0.48\textwidth}
			\centering
			\includegraphics[width=2.7in]{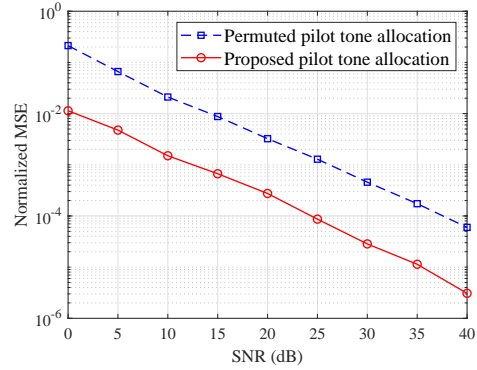}
		\end{minipage}\label{Rel_realCH}
	}\\
	\subfigure[Normalized MSE of different pilot tone allocations versus Rician factor $\kappa$ (dB) with SNR $=20$~dB, $L_1=3$ and $L_2=2$.]{
		\begin{minipage}[b]{0.48\textwidth}
			\centering
			\includegraphics[width=2.7in]{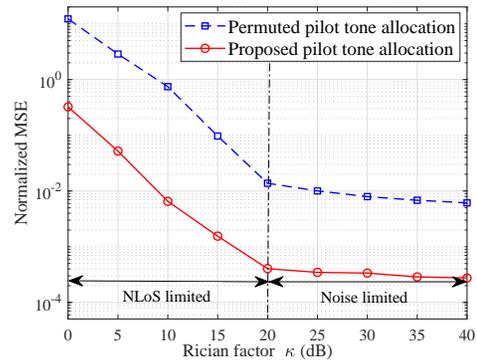}
		\end{minipage}\label{Rel_NLoS_effect}
	}
	\setlength{\abovecaptionskip}{-3pt}
	\caption{Normalized MSE of the SeUCE scheme with the DFT-based IRS reflection pattern.} 
	\vspace{-0.5cm}
\end{figure}
With the equispaced pilot tone allocation and the DFT-based reflection pattern applied to the reference user, we examine the normalized MSE of different pilot tone allocations for
the SeUCE scheme with $L_1=4$ and $L_2=1$ in Fig.~\ref{Rel_realCH}. It is observed that for the SeUCE scheme, the proposed pilot tone allocation design achieves up to $13$~dB SNR gain over the permuted pilot tone allocation benchmark. This can be explained by the fact that 
${{\bm C}}_k$ (given in \eqref{es_mat2}) of the proposed pilot tone allocation design typically has a smaller matrix condition number than that of the permuted benchmark, as verified by a large number of randomly generated ${\bm Q}_1$. 
Note that given any ${\bm Q}_1$, the smaller the matrix condition number of ${{\bm C}}_k$ is, the lower the MSE in \eqref{MSE3} is resulted; and this also holds for the expectation of \eqref{MSE3} over ${\bm Q}_1$, as shown in \eqref{obj_P2.1}.
Therefore, the proposed pilot tone allocation design based on the optimal solution for the system setups with small $N$ and/or $M$ is an effective solution for the general system setups with larger $N$ and/or $M$ for the MSE minimization.

\rev{In Fig.~\ref{Rel_NLoS_effect}, we examine the effect of the multi-path interference in the user-IRS link on the channel estimation performance for the SeUCE scheme, by
showing the normalized MSE versus the Rician factor $\kappa$ (dB) with SNR $=20$ dB, $L_1=3$ and $L_2=2$.
In this case, the channel estimation performance is affected by both the multi-path interference and AWGN.
It is observed that as the Rician factor $\kappa$ increases, the normalized MSE decreases drastically in the range of $\kappa \in [0, 20]$~dB, while it approaches an error floor in the range of $\kappa \in [20, 40]$~dB.
This can be explained by the fact that given SNR $=20$ dB, the channel estimation error is mainly attributed to the NLoS interference as its power is higher than the noise power (i.e., $\kappa<20$~dB); while the channel estimation error mainly results from the noise power when the power of the NLoS components is lower than the noise power (i.e., $\kappa>20$ dB).}
Besides, we observe that for the SeUCE scheme, the proposed pilot tone allocation design always outperforms the permuted pilot tone allocation benchmark, regardless of the NLoS-limited or noise-limited region.

\begin{figure}[!t]
	\centering
	\includegraphics[width=2.7in]{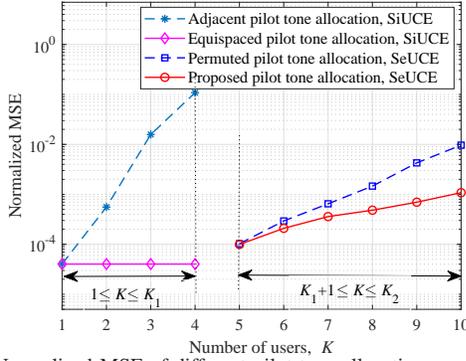}
	\setlength{\abovecaptionskip}{-6pt}
	\caption{Normalized MSE of different pilot tone allocations versus number of users with SNR $=10$~dB, $L_1=4$, and $L_2=1$.}
	\label{Ref_Num_users}
	\vspace{-0.5cm}
\end{figure}

In Fig.~\ref{Ref_Num_users}, we compare the normalized MSE of different pilot tone allocations versus the number of users, $K$, with SNR $=10$~dB, $L_1=4$, and $L_2=1$, assuming the same DFT-based reflection pattern at the IRS.
We observe that when the SiUCE scheme is preferred (i.e., $1\le K \le K_1$), the normalized MSE of the proposed equispaced pilot tone allocation design (see Fig. \ref{peer_pilot}) is invariant to $K$, while that of the adjacent pilot tone allocation benchmark increases dramatically as $K$ increases.
This is expected since the minimum MSE achieved by the SiUCE scheme with the proposed equispaced
pilot tone allocation is irrelevant
to $K$ according to \eqref{MSE_min}.
On the other hand, when the SeUCE scheme is preferred (i.e., $K_1+1\le K \le K_2$), the normalized MSE of both the proposed and permuted pilot tone allocation designs increases with $K$, while the proposed pilot tone allocation design (see Fig. \ref{refer_pilot}) achieves better performance especially for the system with larger $K$.

\section{Conclusions}\label{conlusion}
In this paper, we proposed two efficient uplink channel estimation schemes for different channel setups in the IRS-assisted multi-user OFDMA system. By exploiting the property that all users share the common IRS-AP channel, the proposed SeUCE scheme was shown to be able to achieve more supportable users as compared to the SiUCE scheme, but at the expense of higher channel estimation complexity and some degraded channel estimation performance.
Moreover, for the two proposed channel estimation schemes,
we optimized their corresponding training designs (including user pilot tone allocations and IRS reflection pattern) to minimize the channel estimation error, and derived their fundamental limits on the minimum training overhead and the maximum number of supportable users.
Simulation results
demonstrated the effectiveness of the proposed channel estimation schemes and training designs as compared to heuristic benchmark schemes. Although the proposed schemes apply to the uplink channel estimation at the (multi-antenna) AP, their essential approaches and design methods can be extended to the downlink for each user to estimate its channels from the multi-antenna AP in parallel, by treating each AP antenna/user as an equivalent user/AP antenna in the uplink case.  
\appendix
\section{Proof of Proposition 1}\label{AppendixA}
The objective function in \eqref{obj_P0} can be expanded as 
\begin{align}\label{obj_MSE}
&\sum_{k=1}^K {\mathbb E}\left\{ \left\|  \sqrt{\frac{|{\cal J}_{k}|}{P}} {\bm F}_{k}^\dagger {\bm V}_{k}{\bm \Xi}^\dagger \right\|^{2}_F\right\}\notag\\
=& \sum_{k=1}^K \frac{|{\cal J}_{k}|}{P} \text{tr}\left\{ \left({\bm \Xi}^\dagger \right)^H 
{\mathbb E}\left\{ {\bm V}_{k}^H  ({\bm F}_{k}^\dagger)^H {\bm F}_{k}^\dagger {\bm V}_{k}   \right\} {\bm \Xi}^\dagger	\right\}.
\end{align}
As each ${\bm V}_{k}$ is an AWGN matrix, we have 
\begin{align}\label{exp_V}
&{\mathbb E}\left\{ {\bm V}_{k}^H ({\bm F}_{k}^\dagger)^H {\bm F}_{k}^\dagger  {\bm V}_{k}   \right\}
\notag\\
\hspace{-0.2cm}=&{\mathbb E}\left\{ [{\bm F}_{k}^\dagger{\bm v}_{k}^{(1)},\ldots,{\bm F}_{k}^\dagger{\bm v}_{k}^{(M+1)}]^H   [{\bm F}_{k}^\dagger{\bm v}_{k}^{(1)},\ldots,{\bm F}_{k}^\dagger{\bm v}_{k}^{(M+1)}]   \right\}\notag\\
\hspace{-0.2cm}=&{\mathbb E}\hspace{-0.1cm}\left\{\hspace{-0.1cm}\begin{bmatrix} 
\hspace{-0.2cm} ({\bm v}_{k}^{(1)})^H ({\bm F}_{k}^\dagger)^H {\bm F}_{k}^\dagger {\bm v}_{k}^{(1)} &\hspace{-0.3cm}  \cdots &\hspace{-0.2cm} ({\bm v}_{k}^{(1)})^H  ({\bm F}_{k}^\dagger)^H {\bm F}_{k}^\dagger {\bm v}_{k}^{(M+1)}\hspace{-0.2cm} \\ 
\hspace{-0.2cm} \vdots &\hspace{-0.3cm}  \ddots &\hspace{-0.2cm} \vdots\\ 
\hspace{-0.1cm} ({\bm v}_{k}^{(M+1)})^H \hspace{-0.1cm} ({\bm F}_{k}^\dagger)^H \hspace{-0.1cm} {\bm F}_{k}^\dagger {\bm v}_{k}^{(1)} &\hspace{-0.3cm}  \cdots &\hspace{-0.3cm} ({\bm v}_{k}^{(M+1)})^H ({\bm F}_{k}^\dagger)^H \hspace{-0.1cm} {\bm F}_{k}^\dagger {\bm v}_{k}^{(M+1)}\hspace{-0.5cm}
\end{bmatrix}\hspace{-0.15cm}\right\}
\end{align}
where
\begin{align}
\hspace{-0.2cm}&{\mathbb E}\left\{({\bm v}_{k}^{(t)})^H ({\bm F}_{k}^\dagger)^H {\bm F}_{k}^\dagger {\bm v}_{k}^{(t')}\right\}
\hspace{-0.1cm} =\hspace{-0.1cm} {\mathbb E} \hspace{-0.1cm} \left\{ \text{tr}\left\{ {\bm F}_{k}^\dagger{\bm v}_{k}^{(t')} ({\bm v}_{k}^{(t)})^H ({\bm F}_{k}^\dagger)^H \right\}  \right\} \notag\\
\hspace{-0.2cm}=&\text{tr}\left\{
{\bm F}_{k}^\dagger ~ {\mathbb E}\left\{ {\bm v}_{k}^{(t')} ({\bm v}_{k}^{(t)})^H \right\} ({\bm F}_{k}^\dagger)^H
\right\}\notag\\
=&
\left\{ \begin{gathered}
\sigma^2 \text{tr}\left\{\left({\bm F}_{k}^H {\bm F}_{k}\right)^{-1}\right\},\qquad t=t' \hfill \\
0,\quad\qquad\qquad\qquad\qquad{\rm otherwise} \hfill
\end{gathered}  \right.
\end{align}
since ${\mathbb E}\left\{ {\bm v}_{k}^{(t')} ({\bm v}_{k}^{(t)})^H \right\}=\sigma^2 {\bm I}_{|{\cal J}_{k}|}$ for $t=t'$ and ${\mathbb E}\left\{ {\bm v}_{k}^{(t')} ({\bm v}_{k}^{(t)})^H \right\}= {\bm 0}_{|{\cal J}_{k}|\times |{\cal J}_{k}|}$ for $t \ne t'$.
Accordingly, \eqref{exp_V} can be simplified as
\begin{align}\label{exp_V2}
{\mathbb E}\left\{ {\bm V}_{k}^H ({\bm F}_{k}^\dagger)^H {\bm F}_{k}^\dagger  {\bm V}_{k}   \right\}=
\sigma^2 \text{tr}\left\{\left({\bm F}_{k}^H {\bm F}_{k}\right)^{-1}\right\} {\bm I}_{M+1}.
\end{align}
By substituting \eqref{exp_V2} into \eqref{obj_MSE}, the objective function of problem (P1) is further derived as
\begin{align}\label{obj_MSE2}
&\sum_{k=1}^K {\mathbb E}\left\{ \left\|  \sqrt{\frac{|{\cal J}_{k}|}{P}} {\bm F}_{k}^\dagger {\bm V}_{k}{\bm \Xi}^\dagger \right\|^{2}_F\right\}\notag\\
=&
\sum_{k=1}^K \frac{|{\cal J}_{k}|\sigma^2}{P} 
\text{tr}\left\{\left({\bm F}_{k}^H {\bm F}_{k}\right)^{-1}\right\}
\text{tr}\left\{ \left({\bm \Xi}^\dagger \right)^H  {\bm \Xi}^\dagger	\right\}\notag\\
=& \text{tr}\left\{  \left( {\bm \Xi}  {\bm \Xi}^H \right)^{-1}	\right\}
\sum_{k=1}^K \frac{|{\cal J}_{k}|\sigma^2}{P} 
\text{tr}\left\{\left({\bm F}_{k}^H {\bm F}_{k}\right)^{-1}\right\}.
\end{align}

From \eqref{obj_MSE2}, we see that the optimal joint training design of the IRS reflection pattern and 
	the pilot tone allocation for each user can be decoupled for the SiUCE scheme. 
As such, the optimization
problem (P1) can be equivalently decomposed into two sub-problems as follows.
\begin{align}
\text{(P1.1):}~
& \underset{ \left\{\theta_m^{(t)}\right\} }{\text{min}}
& &\text{tr}\left\{  \left( {\bm \Xi}  {\bm \Xi}^H \right)^{-1}	\right\} \label{obj_P1.1}\\
& \text{~s.t.} & &  |\theta_m^{(t)}|=1, \forall t \in {\cal T}, \forall  m\in {\cal M}.\label{con4_P1.1}\\
\text{(P1.2):}~
& \underset{\left\{\delta_{k,n}\right\}   }{\text{min}}
& &\sum_{k=1}^K \frac{|{\cal J}_{k}|\sigma^2}{P} 
\text{tr}\left\{\left({\bm F}_{k}^H {\bm F}_{k}\right)^{-1}\right\} \label{obj_P1.2}\\
& \text{~s.t.} & &\sum_{k=1}^K \delta_{k,n} \le 1, \forall n \in {\cal N} \label{con2_P1.2}\\
& & &\delta_{k,n} \in\{0,1\}, \forall n \in {\cal N}, \forall k \in {\cal K}\label{con3_P1.2}.
\end{align}

For problem (P1.1), the optimal IRS reflection pattern to minimize the objective function in \eqref{obj_P1.1} should satisfy ${\bm \Xi}  {\bm \Xi}^H =(M+1){\bm I}_{M+1}$ \cite{kay1993fundamentals}, which implies that the optimal IRS reflection pattern ${\bm \Xi}$ is an orthogonal matrix with each entry satisfying the unit-modulus constraint. Moreover, it can be verified that the IRS reflection pattern using the $(M+1)\times (M+1)$ DFT matrix can meet this requirement and thus is an optimal solution to problem (P1.1).
Accordingly, the minimum value of \eqref{obj_P1.1} is given by $\text{tr}\left\{  \left( {\bm \Xi}  {\bm \Xi}^H \right)^{-1}	\right\}=1$.

For problem (P1.2), 
to minimize the objective function in \eqref{obj_P1.2}, we can minimize $\text{tr}\left\{\left({\bm F}_{k}^H {\bm F}_{k}\right)^{-1}\right\}$ for each user $k$.
This optimization problem is equivalent to the MSE minimization problem for traditional multi-user OFDMA systems. According to \cite{Barhumi2003Optimal}, the minimum MSE can be achieved when the pilot tones assigned to each user are equispaced with ${|{\cal J}_{k}|}\ge L$, i.e., ${\cal J}_{k}=\left\{n |n \mod{\frac{N}{|{\cal J}_{k}|}}=j_{k,0}, n \in {\cal N}\right\}$, where 
$\frac{N}{|{\cal J}_{k}|}$ is the spacing of adjacent pilot tones of user $k$ and $j_{k,0}\in \{0,\ldots, \frac{N}{|{\cal J}_{k}|}-1\}$ is the initial pilot tone position, such that it satisfies ${\bm F}_{k}^H {\bm F}_{k}= {\bm F}^H {\bm \Pi}_{{\cal J}_{k}}^T {\bm \Pi}_{{\cal J}_{k}} {\bm F}
=\frac{|{\cal J}_{k}|}{N} {\bm I}_{L} $.
Moreover, due to the disjoint pilot tone allocations for all users, the initial pilot tone position
of each user $j_{k,0}$ should be selected such that ${\cal J}_{k} \bigcap {\cal J}_{k'}=\emptyset $ for $k\ne k'$. Given the above conditions, the minimum value of \eqref{obj_P1.2} is achieved with
\begin{align}
\sum_{k=1}^K \frac{|{\cal J}_{k}|\sigma^2}{P} \text{tr}\left\{\left({\bm F}_{k}^H {\bm F}_{k}\right)^{-1}\right\}
=\frac{\sigma^2NKL}{P}.
\end{align}

Combining the optimal solutions to problems (P1.1) and (P1.2) yields the results in Proposition~1,
thus completing the proof.
\ifCLASSOPTIONcaptionsoff
  \newpage
\fi

\bibliographystyle{IEEEtran}
\bibliography{IRS_OFDM}

\begin{thebibliography}{10}
\providecommand{\url}[1]{#1}
\csname url@samestyle\endcsname
\providecommand{\newblock}{\relax}
\providecommand{\bibinfo}[2]{#2}
\providecommand{\BIBentrySTDinterwordspacing}{\spaceskip=0pt\relax}
\providecommand{\BIBentryALTinterwordstretchfactor}{4}
\providecommand{\BIBentryALTinterwordspacing}{\spaceskip=\fontdimen2\font plus
\BIBentryALTinterwordstretchfactor\fontdimen3\font minus
  \fontdimen4\font\relax}
\providecommand{\BIBforeignlanguage}[2]{{%
\expandafter\ifx\csname l@#1\endcsname\relax
\typeout{** WARNING: IEEEtran.bst: No hyphenation pattern has been}%
\typeout{** loaded for the language `#1'. Using the pattern for}%
\typeout{** the default language instead.}%
\else
\language=\csname l@#1\endcsname
\fi
#2}}
\providecommand{\BIBdecl}{\relax}
\BIBdecl

\bibitem{Andrews2014What}
J.~G. {Andrews}, S.~{Buzzi}, W.~{Choi}, S.~V. {Hanly}, A.~{Lozano}, A.~C.~K.
  {Soong}, and J.~C. {Zhang}, ``What will {5G} be?'' \emph{IEEE J. Sel. Areas
  Commun.}, vol.~32, no.~6, pp. 1065--1082, Jun. 2014.

\bibitem{cui2014coding}
T.~J. Cui, M.~Q. Qi, X.~Wan, J.~Zhao, and Q.~Cheng, ``Coding metamaterials,
  digital metamaterials and programmable metamaterials,'' \emph{L. Sci. \&
  Appl.}, vol.~3, no.~10, pp. e218--e218, Oct. 2014.

\bibitem{liaskos2018new}
C.~Liaskos, S.~Nie, A.~Tsioliaridou, A.~Pitsillides, S.~Ioannidis, and
  I.~Akyildiz, ``A new wireless communication paradigm through
  software-controlled metasurfaces,'' \emph{IEEE Commun. Mag.}, vol.~56, no.~9,
  pp. 162--169, Sept. 2018.

\bibitem{liu2018programmable}
F.~Liu \emph{et~al.}, ``Programmable metasurfaces: State of the art and
  prospects,'' in \emph{Proc. IEEE Int. Symp. Circuits Syst. (ISCAS)},
  Florence, Italy, May 2018, pp. 1--5.

\bibitem{qingqing2019towards}
Q.~Wu and R.~Zhang, ``Towards smart and reconfigurable environment: Intelligent
  reflecting surface aided wireless network,'' \emph{IEEE Commun. Mag.},
  vol.~58, no.~1, pp. 106--112, Jan. 2020.

\bibitem{Renzo2019Smart}
M.~Di~Renzo \emph{et~al.}, ``Smart radio environments empowered by
  reconfigurable {AI} meta-surfaces: An idea whose time has come,''
  \emph{EURASIP J. Wireless Commun. Netw.}, vol. 2019:129, May 2019.

\bibitem{basar2019wireless}
E.~Basar, M.~Di~Renzo, J.~de~Rosny, M.~Debbah, M.-S. Alouini, and R.~Zhang,
  ``Wireless communications through reconfigurable intelligent surfaces,''
  \emph{IEEE Access}, vol.~7, pp. 116\,753--116\,773, Aug. 2019.

\bibitem{Wu2019TWC}
Q.~Wu and R.~Zhang, ``Intelligent reflecting surface enhanced wireless network
  via joint active and passive beamforming,'' \emph{IEEE Trans. Wireless
  Commun.}, vol.~18, no.~11, pp. 5394--5409, Nov. 2019.

\bibitem{wu2019beamforming}
------, ``Beamforming optimization for wireless network aided by intelligent
  reflecting surface with discrete phase shifts,'' \emph{IEEE Trans. Commun.},
  vol.~68, no.~3, pp. 1838--1851, Mar. 2020.

\bibitem{yang2016design}
H.~Yang, X.~Chen, F.~Yang, S.~Xu, X.~Cao, M.~Li, and J.~Gao, ``Design of
  resistor-loaded reflectarray elements for both amplitude and phase control,''
  \emph{IEEE Antennas Wireless Propag. Lett.}, vol.~16, pp. 1159--1162, Nov.
  2016.

\bibitem{zheng2019intelligent}
B.~Zheng and R.~Zhang, ``Intelligent reflecting surface-enhanced {OFDM}:
  Channel estimation and reflection optimization,'' \emph{IEEE Wireless Commun.
  Lett.}, vol.~9, no.~4, pp. 518--522, Apr. 2020.

\bibitem{yang2019intelligent}
Y.~Yang, B.~Zheng, S.~Zhang, and R.~Zhang, ``Intelligent reflecting surface
  meets {OFDM}: Protocol design and rate maximization,'' \emph{IEEE Trans.
  Commun.}, vol.~68, no.~7, pp. 4522--4535, Jul. 2020.

\bibitem{Yang2020IRS}
Y.~{Yang}, S.~{Zhang}, and R.~{Zhang}, ``{IRS}-enhanced {OFDMA}: Joint resource
  allocation and passive beamforming optimization,'' \emph{IEEE Wireless
  Commun. Lett.}, vol.~9, no.~6, pp. 760--764, Jun. 2020.

\bibitem{zhang2019capacity}
S.~Zhang and R.~Zhang, ``Capacity characterization for intelligent reflecting
  surface aided {MIMO} communication,'' \emph{IEEE J. Sel. Areas Commun.},
  vol.~38, no.~8, pp. 1823--1838, Aug. 2020.

\bibitem{Pan2020Multicell}
C.~{Pan}, H.~{Ren}, K.~{Wang}, W.~{Xu}, M.~{Elkashlan}, A.~{Nallanathan}, and
  L.~{Hanzo}, ``Multicell {MIMO} communications relying on intelligent
  reflecting surfaces,'' \emph{IEEE Trans. Wireless Commun.}, vol.~19, no.~8,
  pp. 5218--5233, Aug. 2020.

\bibitem{Zheng2020IRSNOMA}
B.~{Zheng}, Q.~{Wu}, and R.~{Zhang}, ``Intelligent reflecting surface-assisted
  multiple access with user pairing: {NOMA or OMA}?'' \emph{IEEE Commun.
  Lett.}, vol.~24, no.~4, pp. 753--757, Apr. 2020.

\bibitem{yanggang2019intelligent}
G.~{Yang}, X.~{Xu}, and Y.~{Liang}, ``Intelligent reflecting surface assisted
  non-orthogonal multiple access,'' in \emph{Proc. IEEE Wireless Commun. Netw.
  Conf. (WCNC)}, Seoul, Korea (South), May 2020, pp. 1--6.

\bibitem{Wu2020Weighted}
Q.~{Wu} and R.~{Zhang}, ``Weighted sum power maximization for intelligent
  reflecting surface aided {SWIPT},'' \emph{IEEE Wireless Commun. Lett.},
  vol.~9, no.~5, pp. 586--590, May 2020.

\bibitem{Wu2020Weighted2}
------, ``Joint active and passive beamforming optimization for intelligent
  reflecting surface assisted {SWIPT} under {QoS} constraints,'' \emph{IEEE J.
  Sel. Areas Commun}, vol.~38, no.~8, pp. 1735--1748, Aug. 2020.

\bibitem{Pan2020Intelligent}
C.~{Pan}, H.~{Ren}, K.~{Wang}, M.~{Elkashlan}, A.~{Nallanathan}, J.~{Wang}, and
  L.~{Hanzo}, ``Intelligent reflecting surface aided {MIMO} broadcasting for
  simultaneous wireless information and power transfer,'' \emph{IEEE J. Sel.
  Areas Commun.}, vol.~38, no.~8, pp. 1719--1734, Aug. 2020.

\bibitem{Cui2019Secure}
M.~{Cui}, G.~{Zhang}, and R.~{Zhang}, ``Secure wireless communication via
  intelligent reflecting surface,'' \emph{IEEE Wireless Commun. Lett.}, vol.~8,
  no.~5, pp. 1410--1414, Oct. 2019.

\bibitem{Guan2020Intelligent}
X.~{Guan}, Q.~{Wu}, and R.~{Zhang}, ``Intelligent reflecting surface assisted
  secrecy communication: Is artificial noise helpful or not?'' \emph{IEEE
  Wireless Commun. Lett.}, vol.~9, no.~6, pp. 778--782, Jun. 2020.

\bibitem{chen2019intelligent}
J.~Chen, Y.-C. Liang, Y.~Pei, and H.~Guo, ``Intelligent reflecting surface: A
  programmable wireless environment for physical layer security,'' \emph{IEEE
  Access}, vol.~7, pp. 82\,599--82\,612, Jun. 2019.

\bibitem{Guan2020Joint}
X.~{Guan}, Q.~{Wu}, and R.~{Zhang}, ``Joint power control and passive
  beamforming in {IRS}-assisted spectrum sharing,'' \emph{IEEE Commun. Lett.},
  vol.~24, no.~7, pp. 1553--1557, Jul. 2020.

\bibitem{jensen2019optimal}
T.~L. Jensen and E.~De~Carvalho, ``On optimal channel estimation scheme for
  intelligent reflecting surfaces based on a minimum variance unbiased
  estimator,'' in \emph{Proc. IEEE Int. Conf. Acoust., Speech, Signal Process.
  (ICASSP)}, Barcelona, Spain, May 2020, pp. 5000--5004.

\bibitem{you2019intelligent}
C.~You, B.~Zheng, and R.~Zhang, ``Intelligent reflecting surface with discrete
  phase shifts: Channel estimation and passive beamforming,'' in \emph{Proc.
  IEEE Int. Conf. Commun. (ICC)}, Dublin, Ireland, Jun. 2020, pp. 1--6.

\bibitem{you2019progressive}
------, ``Channel estimation and passive beamforming for intelligent reflecting
  surface: Discrete phase shift and progressive refinement,'' \emph{IEEE J.
  Sel. Areas Commun}, doi: 10.1109/JSAC.2020.3007056, Jul. 2020.

\bibitem{zhou2020framework}
G.~Zhou, C.~Pan, H.~Ren, K.~Wang, and A.~Nallanathan, ``A framework of robust
  transmission design for {IRS}-aided {MISO} communications with imperfect
  cascaded channels,'' \emph{IEEE Trans. Signal Process.}, doi:
  10.1109/TSP.2020.3019666, Aug. 2020.

\bibitem{zhao2019intelligent}
M.-M. Zhao, Q.~Wu, M.-J. Zhao, and R.~Zhang, ``Intelligent reflecting surface
  enhanced wireless network: Two-timescale beamforming optimization,''
  \emph{arXiv preprint arXiv:1912.01818}, 2019.

\bibitem{he2019cascaded}
Z.-Q. He and X.~Yuan, ``Cascaded channel estimation for large intelligent
  metasurface assisted massive {MIMO},'' \emph{IEEE Wireless Commun. Lett.},
  vol.~9, no.~2, pp. 210--214, Feb. 2020.

\bibitem{mirza2019channel}
J.~Mirza and B.~Ali, ``Channel estimation method and phase shift design for
  reconfigurable intelligent surface assisted {MIMO} networks,'' \emph{arXiv
  preprint arXiv:1912.10671}, 2019.

\bibitem{wang2019compressed}
P.~Wang, J.~Fang, H.~Duan, and H.~Li, ``Compressed channel estimation for
  intelligent reflecting surface-assisted millimeter wave systems,'' \emph{IEEE
  Signal Process. Lett.}, vol.~27, pp. 905--909, May 2020.

\bibitem{chen2019channel}
J.~Chen, Y.-C. Liang, H.~V. Cheng, and W.~Yu, ``Channel estimation for
  reconfigurable intelligent surface aided multi-user {MIMO} systems,''
  \emph{arXiv preprint arXiv:1912.03619}, 2019.

\bibitem{liu2019matrix}
H.~Liu, X.~Yuan, and Y.~Jun, ``Matrix-calibration-based cascaded channel
  estimation for reconfigurable intelligent surface assisted multiuser
  {MIMO},'' \emph{IEEE J. Sel. Areas Commun}, doi: 10.1109/JSAC.2020.3007057,
  Jul. 2020.

\bibitem{wang2019channel}
Z.~{Wang}, L.~{Liu}, and S.~{Cui}, ``Channel estimation for intelligent
  reflecting surface assisted multiuser communications: Framework, algorithms,
  and analysis,'' \emph{IEEE Trans. Wireless Commun.}, doi:
  10.1109/TWC.2020.3004330, Jun. 2020.

\bibitem{wei2020parallel}
L.~Wei, C.~Huang, G.~C. Alexandropoulos, and C.~Yuen, ``Parallel factor
  decomposition channel estimation in {RIS}-assisted multi-user {MISO}
  communication,'' in \emph{Proc. IEEE Sensor Array Multichannel Signal
  Process. Workshop (SAM)}, Hangzhou, China, China, Jun. 2020, pp. 1--5.

\bibitem{kay1993fundamentals}
S.~M. Kay, \emph{Fundamentals of statistical signal processing}.\hskip 1em plus
  0.5em minus 0.4em\relax Prentice Hall PTR, 1993.

\bibitem{Barhumi2003Optimal}
I.~{Barhumi}, G.~{Leus}, and M.~{Moonen}, ``Optimal training design for {MIMO
  OFDM} systems in mobile wireless channels,'' \emph{IEEE Trans. Signal
  Process.}, vol.~51, no.~6, pp. 1615--1624, Jun. 2003.

\end{thebibliography}

\begin{IEEEbiography}[{\includegraphics[width=1in,height=1.25in,clip,keepaspectratio]{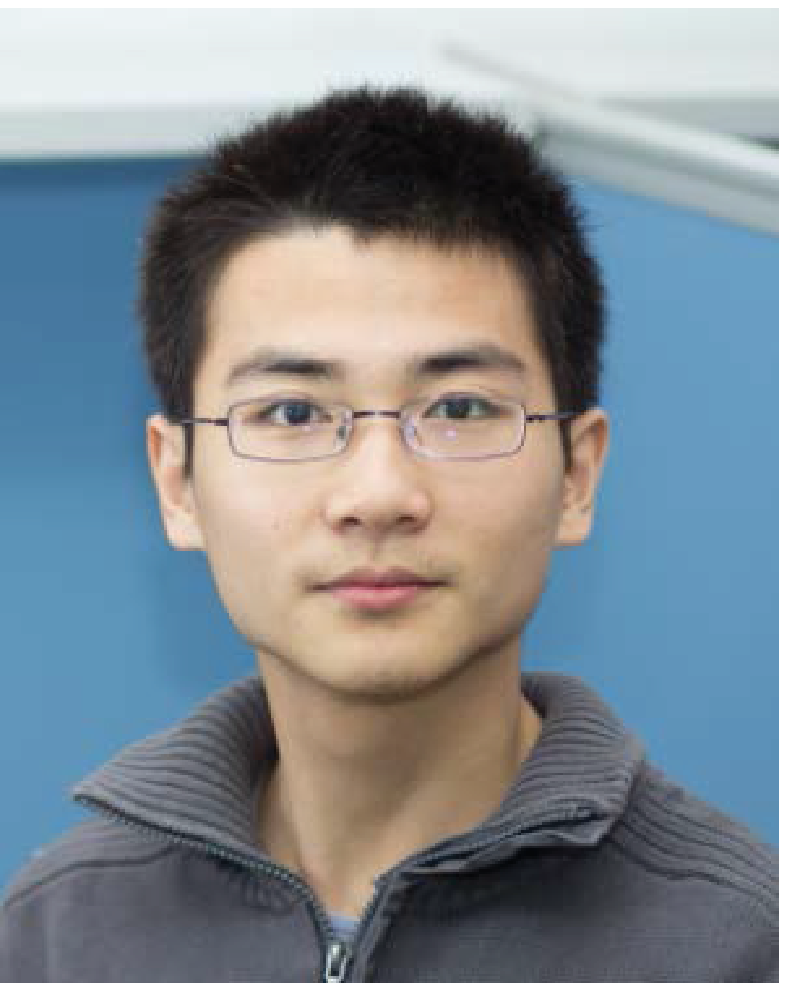}}]{Beixiong Zheng}
	(M'18) received the B.S. and Ph.D.
	degrees from the South China University of Technology, Guangzhou, China, in 2013 and 2018, respectively. 
	He is currently a Research Fellow with the Department of Electrical and Computer Engineering, National University of Singapore.
	His recent research interests include intelligent reflecting surface (IRS), index modulation (IM), and non-orthogonal multiple access (NOMA). 
	
	From 2015 to 2016, he was a Visiting Student Research Collaborator with Columbia University, New York, NY, USA. He was a recipient of
	the Best Paper Award from the IEEE International Conference on Computing, Networking and Communications in 2016, 
	the Best Ph.D. Thesis Award from China Education Society of Electronics in 2018,
	and the Outstanding Reviewer of Physical Communication in 2019.
\end{IEEEbiography}

\begin{IEEEbiography}[{\includegraphics[width=1in,height=1.25in,clip,keepaspectratio]{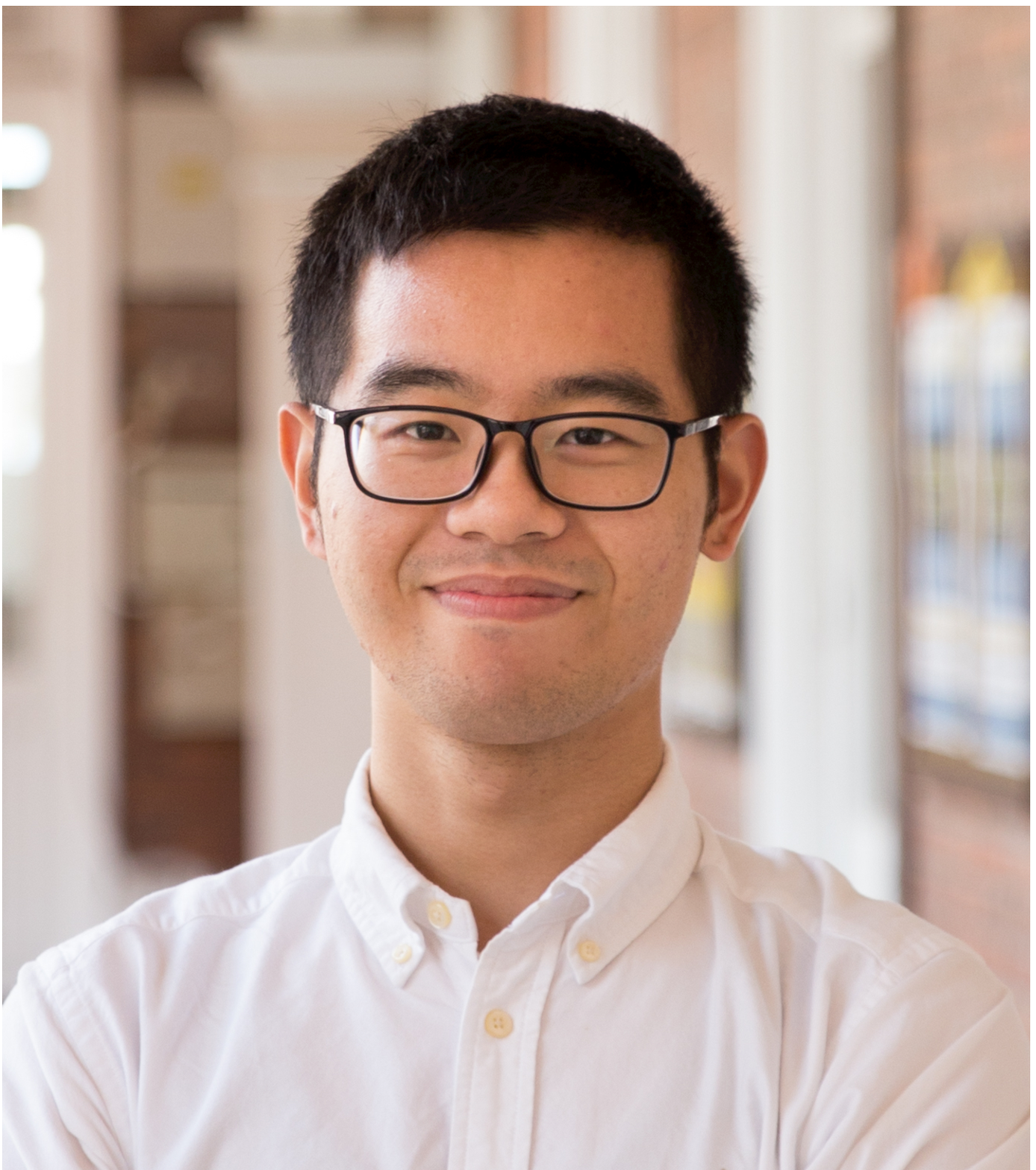}}]{Changsheng You}(M'19) received his B.Eng. degree in 2014 from University of Science and Technology of China in electronic engineering and information science, and  Ph.D. degree in 2018 from The University of Hong Kong in electrical and electronic engineering. He is currently a Research Fellow with the Department of Electrical and Computer Engineering, National University of Singapore. His research interests include intelligent reflecting surface, UAV communications, edge learning, mobile-edge computing, wireless power transfer, and convex optimization.
	
Dr. You received the IEEE Communications Society Asia-Pacific Region Outstanding Paper Award in 2019 and the Exemplary Reviewer of the IEEE TRANSACTIONS ON COMMUNICATIONS and IEEE TRANSACTIONS ON WIRELESS COMMUNICATIONS. He was a recipient of the Best Ph.D. Thesis Award of The University of Hong Kong in 2019.
\end{IEEEbiography}

\begin{IEEEbiography}[{\includegraphics[width=1in,height=1.25in,clip,keepaspectratio]{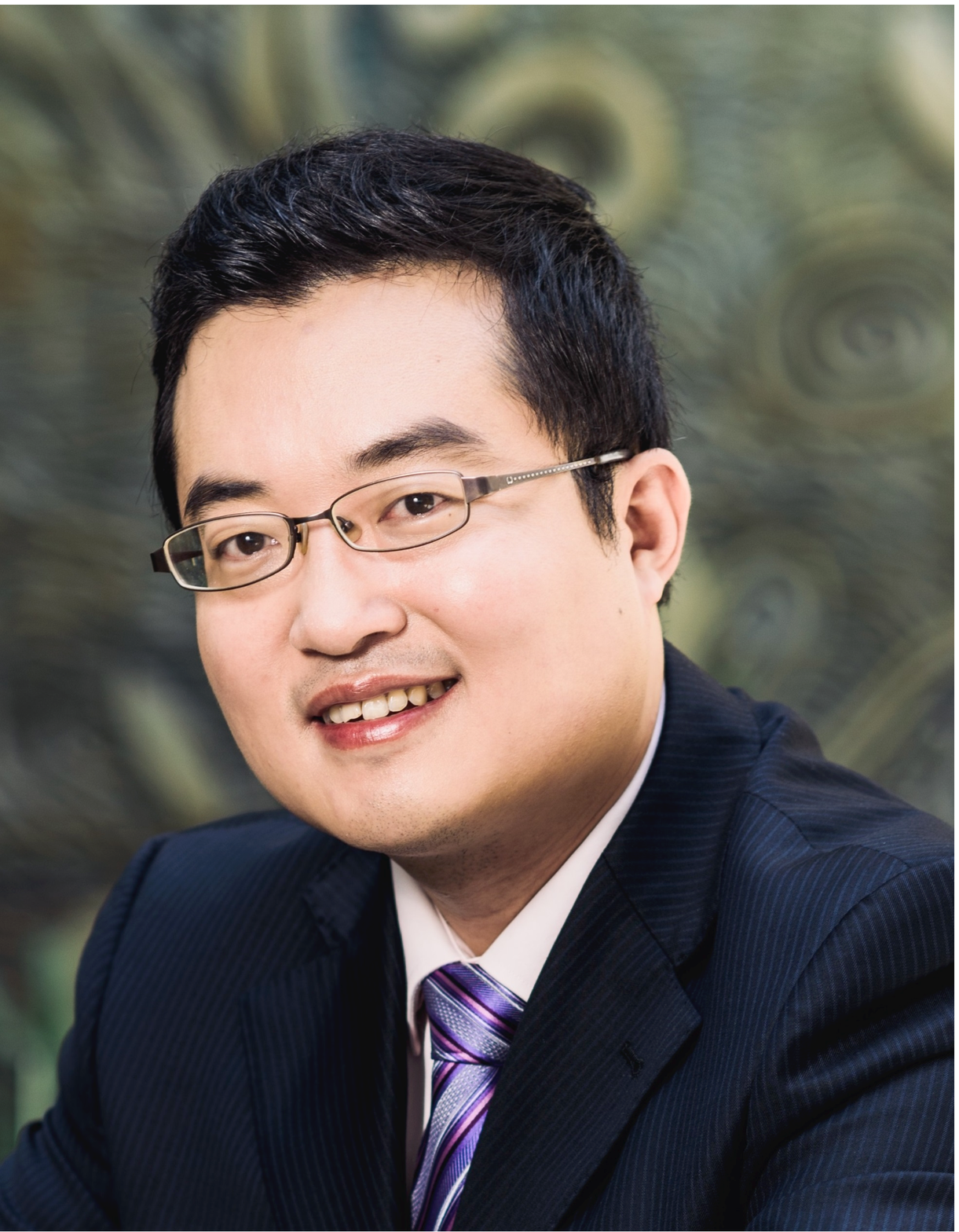}}] {Rui Zhang} (S'00-M'07-SM'15-F'17) received the B.Eng. (first-class Hons.) and M.Eng. degrees from the National University of Singapore, Singapore, and the Ph.D. degree from the Stanford University, Stanford, CA, USA, all in electrical engineering.
	
	From 2007 to 2010, he worked at the Institute for Infocomm Research, ASTAR, Singapore. Since 2010, he has been working with the National University of Singapore, where he is now a Professor in the Department of Electrical and Computer Engineering. He has published over 200 journal papers and over 180 conference papers. He has been listed as a Highly Cited Researcher by Thomson Reuters/Clarivate Analytics since 2015. His current research interests include UAV/satellite communications, wireless power transfer, reconfigurable MIMO, and optimization methods.     
	
	He was the recipient of the 6th IEEE Communications Society Asia-Pacific Region Best Young Researcher Award in 2011, and the Young Researcher Award of National University of Singapore in 2015. He was the co-recipient of the IEEE Marconi Prize Paper Award in Wireless Communications in 2015 and 2020, the IEEE Communications Society Asia-Pacific Region Best Paper Award in 2016, the IEEE Signal Processing Society Best Paper Award in 2016, the IEEE Communications Society Heinrich Hertz Prize Paper Award in 2017 and 2020, the IEEE Signal Processing Society Donald G. Fink Overview Paper Award in 2017, and the IEEE Technical Committee on Green Communications \& Computing (TCGCC) Best Journal Paper Award in 2017. His co-authored paper received the IEEE Signal Processing Society Young Author Best Paper Award in 2017. He served for over 30 international conferences as the TPC co-chair or an organizing committee member, and as the guest editor for 3 special issues in the IEEE JOURNAL OF SELECTED TOPICS IN SIGNAL PROCESSING and the IEEE JOURNAL ON SELECTED AREAS IN COMMUNICATIONS. He was an elected member of the IEEE Signal Processing Society SPCOM Technical Committee from 2012 to 2017 and SAM Technical Committee from 2013 to 2015, and served as the Vice Chair of the IEEE Communications Society Asia-Pacific Board Technical Affairs Committee from 2014 to 2015. He served as an Editor for the IEEE TRANSACTIONS ON WIRELESS COMMUNICATIONS from 2012 to 2016, the IEEE JOURNAL ON SELECTED AREAS IN COMMUNICATIONS: Green Communications and Networking Series from 2015 to 2016, the IEEE TRANSACTIONS ON SIGNAL PROCESSING from 2013 to 2017, and the IEEE TRANSACTIONS ON GREEN COMMUNICATIONS AND NETWORKING from 2016 to 2020. He is now an Editor for the IEEE TRANSACTIONS ON COMMUNICATIONS. He serves as a member of the Steering Committee of the IEEE Wireless Communications Letters. He is a Distinguished Lecturer of IEEE Signal Processing Society and IEEE Communications Society. 
\end{IEEEbiography}

\end{document}